\documentclass[aps,prd,preprintnumbers,superscriptaddress,showpacs,showkeys,floatfix,onecolumn,nofootinbib]{revtex4-2}

\usepackage{amsmath,amssymb,amsfonts,dcolumn}
\usepackage{graphicx}

\usepackage[dvipsnames]{xcolor}

\usepackage{hyperref}

\usepackage{braket} 
\usepackage{slashed} 

\usepackage{csquotes}

\usepackage{upgreek} 

\usepackage{enumitem} 

\usepackage{booktabs,tabulary}
\usepackage{multirow}

\newcommand{\beq}{\begin{equation}}
\newcommand{\eeq}{\end{equation}}

\newcommand{\LN}[1]{\textsc{#1}} 

\newcommand{\Lat}[1]{\textit{#1}} 

\newcommand{\tior}{\mathrm{T}}

\newcommand{\eps}{\epsilon}

\newcommand{\ex}{\mathrm{e}}

\newcommand{\iu}{\mathrm{i}}

\renewcommand{\vec}[1]{\boldsymbol{#1}} 

\newcommand{\dff}{\mathrm{d}}
\newcommand{\dfx}{\mathrm{d}^4x\,}

\newcommand{\diag}{\text{diag}}
\newcommand{\Tr}{\text{Tr}}

\newcommand{\discq}{\text{Disc}_{q^2}}
\newcommand{\discs}{\text{Disc}_s}

\newcommand{\Order}{\mathcal{O}}

\renewcommand{\Re}{\text{Re}}

\newcommand{\sgn}{\text{sgn}}

\newcommand{\Cov}{\mathrm{Cov}}

\newcommand{\ellp}{\ell^{\prime-}}
\newcommand{\ellpbar}{\ell^{\prime+}}

\newcommand{\mB}{m_B}
\newcommand{\mBOne}{m_{B_1}}
\newcommand{\mBStar}{m_{B^*}}

\newcommand{\Momega}{M_\omega}
\newcommand{\Mrho}{M_\rho}

\newcommand{\Mpi}{M_\pi}

\newcommand{\Op}[1]{\mathcal{O}_{#1}^{ub\ell\nu}}
\newcommand{\Wilson}[1]{\mathcal{C}_{#1}^{ub\ell\nu}}

\newcommand{\FC}{G_\mathrm{F}}

\newcommand{\weak}{\text{W}}
\newcommand{\had}{\text{H}}
\newcommand{\elmag}{\text{EM}}

\newcommand{\FSR}{\text{FSR}}

\newcommand{\ps}{P}

\newcommand{\M}{\mathcal{M}}

\newcommand{\F}{\mathcal{F}}

\newcommand{\OPE}{\text{OPE}}

\newcommand{\Q}{\mathcal{Q}}

\newcommand{\BW}{\text{BW}}

\renewcommand{\hom}{\text{hom.}}
\newcommand{\inhom}{\text{inhom.}}

\newcommand{\proj}{\mathcal{P}}

\newcommand{\LabelA}{\mathcal{A}}
\newcommand{\LabelB}{\mathcal{B}}
\newcommand{\LabelC}{\mathcal{C}}
\newcommand{\LabelD}{\mathcal{D}}

\newcommand{\BR}{\mathcal{B}}
\newcommand{\FB}{\text{FB}}

\newcommand{\CP}{\mathrm{CP}}

\providecommand{\MeV}{\,\text{MeV}}
\providecommand{\GeV}{\,\text{GeV}}

\newcommand{\perc}{\%}

\begin{document}

\renewcommand{\figureautorefname}{Fig.}
\renewcommand{\tableautorefname}{Table}
\renewcommand{\chapterautorefname}{Ch.}
\renewcommand{\sectionautorefname}{Sec.}
\renewcommand{\subsectionautorefname}{Sec.}
\renewcommand{\appendixautorefname}{App.}
\def\equationautorefname~#1\null{Eq.~(#1)\null}

\preprint{IPPP/22/63, TUM-HEP 1421/22}
\title{Dispersion relations for \boldmath{$B^- \to \ell^- \bar{\nu}_\ell \ellp \ellpbar$} form factors}

\author{Stephan K\"urten}
\email{stephan.kuerten91@gmail.com}
\affiliation{Physik Department (T31), Technische Universit\"at M\"unchen, 85748 Garching, Germany}

\author{Marvin Zanke}
\email{zanke@hiskp.uni-bonn.de}
\affiliation{Helmholtz-Institut für Strahlen- und Kernphysik (Theorie) and\\ Bethe Center for Theoretical Physics, Universität Bonn, 53115 Bonn, Germany}

\author{Bastian Kubis}
\email{kubis@hiskp.uni-bonn.de}
\affiliation{Helmholtz-Institut für Strahlen- und Kernphysik (Theorie) and\\ Bethe Center for Theoretical Physics, Universität Bonn, 53115 Bonn, Germany}

\author{Danny van Dyk}
\email{danny.van.dyk@gmail.com}
\affiliation{Physik Department (T31), Technische Universit\"at M\"unchen, 85748 Garching, Germany}
\affiliation{Institute for Particle Physics Phenomenology and\\ Department of Physics, Durham University, Durham DH1 3LE, United Kingdom}

\begin{abstract}
Using dispersive methods, we study the $B \to \gamma^*$ form factors underlying the decay $B^- \to \ell^- \bar{\nu}_\ell \ellp \ellpbar$.
We discuss the ambiguity that arises from a separation of the full $B^- \to \ell^- \bar{\nu}_\ell \ellp \ellpbar$ amplitude into a hadronic tensor and a final-state-radiation piece, including effects from nonvanishing lepton masses.
For the eligibility of a dispersive treatment, we propose a decomposition of the hadronic part that leads to four form factors that are free of kinematic singularities.
By establishing a set of dispersion relations, we then relate the $B \to \gamma^*$ form factors to the well-known $B \to V$, $V=\omega(782),\rho(770)$, analogs.
Using the combination of a series expansion in a conformal variable and a vector-meson-dominance ansatz to parameterize the $B \to \gamma^*$ form factors, we infer the values of the associated unknown parameters from the available input on $B \to V$.
The phenomenological application of our formalism includes the determination of the branching ratios and forward--backward asymmetries of the process $B^- \to \ell^- \bar{\nu}_\ell \ellp \ellpbar$.
\end{abstract}


\maketitle

\section{Introduction}
\label{sec:introduction}
The radiative leptonic decay $B^- \to \ell^- \bar{\nu}_\ell \gamma$ is widely considered to be the best source of information on the leading-twist $B$-meson light-cone distribution amplitude (LCDA) by elucidating the inner structure of the $B$ meson~\cite{Beneke:2011nf,Wang:2016qii,Beneke:2018wjp}.
However, measurements of this decay are likely only possible at the ongoing Belle~II experiment and not at the LHC experiments, primarily the LHCb.
This precludes leveraging the upcoming large datasets at the LHC, which will become available from run~3 onwards.
The four-lepton decay of the $B$ meson, $B^- \to \ell^- \bar{\nu}_\ell \ellp \ellpbar$, with $\ell' \neq \ell$, $\ell^{(\prime)} = e, \mu$, has been identified as a suitable candidate for studies at both Belle~II and the LHC experiments.
This decay has been studied to some extent in the literature, with a variety of models for the relevant $B \to \gamma^{*}$ form factors~\cite{Beneke:2021rjf,Ivanov:2021jsr,Albrecht:2019zul,Wang:2021yrr}.
However, its usefulness to extract $B$-meson LCDA parameters is hampered by the need for a description of a virtual photon in the timelike region, which requires careful treatment.

We propose a dispersive approach for $B \to \gamma^*$, which is based on the fundamental principles of analyticity and unitarity.
Dispersive analyses in the timelike region are commonly done for low-energy processes, such as the pion vector form factor; see, for instance, Ref.~\cite{Colangelo:2018mtw} and references therein.
Here, we apply methods originally developed for these processes to hadronic transition form factors of $B$ mesons.
For future analyses, our approach has the potential to enable the transfer of information from the region of timelike photon momentum to the spacelike region, where the sensitivity to the LCDA parameters is less affected by soft interactions~\cite{Beneke:2018wjp}.
We relate the isoscalar and isovector components of the $B \to \gamma^*$ transition inherent to the hadronic part of the amplitude through $B^- \to \ell^- \bar{\nu}_\ell \gamma^* (\to \ellp \ellpbar)$ to available input on $B \to \omega \equiv \omega(782)$ and $B \to \rho \equiv \rho(770)$~\cite{Bharucha:2015bzk} via a set of dispersion relations in the photon momentum.
Although we use a vector-meson-dominance (VMD) ansatz in this work, our results provide the groundwork for more sophisticated future analyses.
Using dispersion relations requires the form factors to be free of kinematic singularities.
We modify the well-known \LN{Bardeen}--\LN{Tung}--\LN{Tarrach} (BTT)~\cite{Bardeen:1968ebo,Tarrach:1975tu} procedure, which has not been designed for hadronic form factors in weak transitions, to obtain such a set of form factors.
At this, we face a problem: the separation of the amplitude into a hadronic term---containing the nonperturbative dynamics of the process---and a final-state-radiation (FSR) term turns out to be ambiguous; the two terms are not individually gauge invariant but only their sum is.
A further issue is the lack of definite angular-momentum and parity quantum numbers of the form factors.
Our modification to the BTT procedure addresses this issue, and we take special care not to spoil the singularity-free structure.

To ensure a consistent treatment of lepton-mass effects, we work with nonzero lepton masses throughout our analysis; taking the limit $m_{\ell^{(\prime)}} \to 0$ remains possible.
While the considerations in this article are mostly restricted to the decay of a negatively charged $B$ meson, the decay of a positively charged $B$ meson can be calculated in complete analogy, with some minor adjustments to the formulae given here and completely equivalent numerical results.

The outline of this article is as follows: in \autoref{sec:WET}, we introduce the Lagrangian of the weak effective theory (WET) that describes semileptonic $b\to u\ell\bar\nu$ transitions.
The amplitude for $B^- \to \ell^- \bar{\nu}_\ell \gamma^* (\to \ellp \ellpbar)$ and its decomposition into a hadronic tensor and an FSR piece is discussed in \autoref{sec:hadronic_tensor}.
Using our modified BTT procedure, the hadronic tensor is then parameterized in terms of four form factors that are free of kinematic singularities in \autoref{sec:form_factors}, where the ambiguity arising from the separation of the full amplitude is a subject of special attention.
In \autoref{sec:dispersion_relations}, we establish a set of dispersion relations that relate the $B^- \to \gamma^*$ transition inherent to the hadronic part of the amplitude to available input on $B^- \to V$ form factors, $V=\omega, \rho$, and provide predictions for the $B^-\to \gamma^*$ form factors.
Using these predictions, we present numerical results for the branching ratios and forward--backward (FB) asymmetries of the process $B^- \to \ell^- \bar{\nu}_\ell \ellp \ellpbar$ in \autoref{sec:pheno}.
We conclude and give a brief outlook in \autoref{sec:summary_outlook}.
Some supplementary material is outsourced to Apps.~\ref{app:inhomogeneities}--\ref{app:constants}.

\section{Weak effective theory}
\label{sec:WET}
At the energy scale of the $B$ meson, the standard model's (SM's) flavor-changing processes are conveniently described within an effective field theory~\cite{Aebischer:2017gaw,Jenkins:2017jig}.
The leading terms in this theory arise at mass dimension six, with higher-dimensional operators being suppressed by at least $\mB^2/M_W^2 \approx 0.4\perc$.
Moreover, such an effective field theory allows us to transparently include potential effects beyond the SM as long as new matter fields and mediators live above the scale of electroweak symmetry breaking.
For $b \to u \ell\bar{\nu}_\ell$ transitions in particular, we use the effective Lagrangian
\beq
  \mathcal{L}_\text{WET}^{ub\ell\nu} 
  = \frac{4 \FC}{\sqrt{2}} V_{ub} \sum_{i} \Wilson{i} \Op{i} + \text{h.c.},
\eeq
where $\FC$ is the \LN{Fermi} constant as measured in muon decays, $V_{ub}$ is the \LN{Cabibbo}--\LN{Kobayashi}--\LN{Maskawa} (CKM) matrix element for the $b \rightarrow u$ transition, and $\Wilson{i} \equiv \Wilson{i}(\mu)$ are the so-called \LN{Wilson} coefficients at the scale $\mu$ that multiply the local field operators $\Op{i} \equiv \Op{i}(x)$.
A convenient basis of operators up to dimension six and with only left-handed neutrinos is given by
\begin{align}
    \Op{V,L(R)}
        &= \big[\bar{u}(x) \gamma^\mu P_{L(R)} b(x)\big] \big[\bar{\ell}(x) \gamma_\mu P_L \nu_\ell(x)\big], 
        &
    \Op{S,L(R)}
        &= \big[\bar{u}(x) P_{L(R)} b(x)\big] \big[\bar{\ell}(x) P_L \nu_\ell(x)\big],     
    \notag \\
    \Op{T}
        &= \big[\bar{u}(x) \sigma^{\mu\nu} b(x)\big] \big[\bar{\ell}(x) \sigma_{\mu\nu} P_L \nu_\ell(x)\big],
\end{align}
where, in the SM, $\Wilson{V,L}\rvert_\text{SM} = 1 + \Order(\alpha_e)$ and $\Wilson{i}\rvert_\text{SM} = 0$ for all other corresponding \LN{Wilson} coefficients.
Here, $P_{L/R} = (1 \mp \gamma_5)/2$ are the projection operators onto the left- and right-chiral components and $\alpha_e = e^2/(4\pi)$ is the fine-structure constant.
To leading order in the electromagnetic (EM) interaction, matrix elements of the above operators factorize into matrix elements of a purely hadronic and a purely leptonic current.
In this work, we limit ourselves to the SM operator $\Op{V,L}$ and---to a lesser extent---the scalar operator $\Op{S,L}$.

\section{\boldmath{Hadronic tensor}}
\label{sec:hadronic_tensor}
\begin{figure}[t]
    \centering
    \includegraphics[scale=0.875]{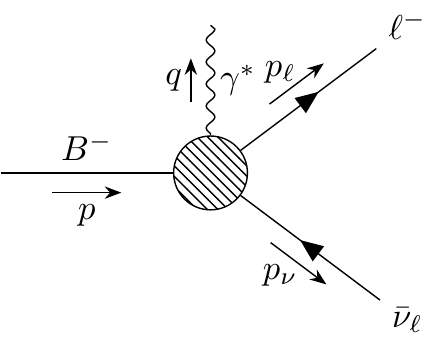}
    \qquad
    \qquad
    \qquad
    \qquad
    \includegraphics[scale=0.875]{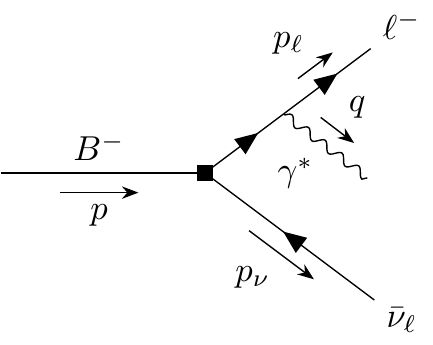}
    \caption{The diagrams contributing to the decay $B^- \to \ell^- \bar{\nu}_\ell \gamma^*$ at dimension six in the WET on the hadronic level: pole and cut contributions of $T_\had^{\mu \nu}(k,q)$, \Lat{e.g.}, from the intermediate states $B$ in $k^2$ or $\pi \pi$ in $q^2$ (\textit{left}) and emission from the charged final-state lepton in $T_\FSR^\mu(p_\ell,p_\nu,q)$ (\textit{right}).
    The hadronic tensor $T_\had^{\mu \nu}(k,q)$ and FSR tensor $T_\FSR^\mu(p_\ell,p_\nu,q)$ are defined in Eqs.~\eqref{eq:definition_hadronic_tensor} and \eqref{eq:definition_FSR_tensor}, respectively.
    Note that an effective four-particle vertex is discarded here, since it contributes at dimension eight in the WET.}
    \label{fig:B_to_l_nu_gamma_hadronic_diagrams}
\end{figure}

We study the decay $B^-(p) \to \ell^-(p_\ell) \bar{\nu}_\ell(p_\nu) \gamma^*(q)$, $k = p_\ell + p_\nu$, whose amplitude in the SM reads~\cite{Beneke:2011nf}
\beq\label{eq:M_B_to_l_nu_gamma}
	\M(B^- \to \ell^- \bar{\nu}_\ell \gamma^*) 
	= \frac{4 \FC V_{ub}}{\sqrt{2}} \braket{\ell^- \bar{\nu}_\ell \gamma^* |\Op{V,L}| B^-}
\eeq
up to corrections of $\Order(\alpha_e)$.
It is convenient to write the WET operator in terms of the leptonic and hadronic weak currents $J_\weak^\nu(x) = \bar{\ell}(x) \gamma^\nu (1-\gamma_5) \nu_\ell(x)$ and $J_H^\nu(x) = \bar{u}(x) \gamma^\nu (1-\gamma_5) b(x)$ according to
\beq
    \Op{V,L} 
    = \frac{1}{4} {J_\had}_\nu(0) J_\weak^\nu(0).
\eeq
At the level of the WET, there are two possible diagrammatic ways for the emission of the (virtual) photon: either from the constituents of the $B$ meson or from the charged final-state lepton; the respective diagrams are shown in~\autoref{fig:B_to_l_nu_gamma_hadronic_diagrams}.

At leading order in the EM coupling, the hadronic matrix element on the right-hand side of \autoref{eq:M_B_to_l_nu_gamma} can be written as
\begin{align}\label{eq:matrix_element_B_to_l_nu_gamma}
	\braket{\ell^- \bar{\nu}_\ell \gamma^* | {J_\had}_\nu(0) J_\weak^\nu(0) | B^-}
	&= e \eps_\mu^* \Big[ \braket{\ell^- \bar{\nu}_\ell | {J_\weak}_\nu(0) | 0} \int\dfx \ex^{\iu q x} \braket{0 | \tior\{J_\elmag^\mu(x) J_\had^\nu(0)\} | B^-} \notag \\
	&\qquad \quad + \braket{0 | {J_\had}_\nu(0) | B^-} \int\dfx \ex^{\iu q x} \braket{\ell^- \bar{\nu}_\ell | \tior\{J_\elmag^\mu(x) J_\weak^\nu(0)\} | 0} \Big] \notag \\
	&= e \eps_\mu^* \Big[ Q_B L_\nu T_\had^{\mu \nu}(k,q) - \iu f_B p_\nu \int\dfx \ex^{\iu q x} \braket{\ell^- \bar{\nu}_\ell | \tior\{J_\elmag^\mu(x) J_\weak^\nu(0)\} | 0} \Big] \notag \\
	&= e \eps_\mu^* \big[ Q_B L_\nu T_\had^{\mu \nu}(k,q) + Q_\ell T_\FSR^\mu(p_\ell,p_\nu,q) \big],
\end{align}
where $e$ is the elementary charge and $\eps_\mu^* \equiv \eps_\mu^*(q; \lambda)$ the polarization vector of the outgoing photon with momentum $q$ and polarization $\lambda$.
Furthermore, $f_B$ is the decay constant of the $B$-meson, $\braket{0 | \bar{u}(0) \gamma^\nu \gamma_5 b(0) | B^-} = \iu f_B p^\nu$, and 
\beq
    J_\elmag^\mu(x) 
    = \bar{q}(x) \Q \gamma^\mu q(x) + \sum_\ell Q_\ell \bar{\ell}(x) \gamma^\mu \ell(x)
\eeq
the EM current, with $q(x) = (u(x),d(x),s(x),c(x),b(x))^\intercal$, $\Q = \diag[2/3,-1/3,-1/3, 2/3, -1/3]$ the quark charge matrix, and $Q_B=-1=Q_\ell$ the charge of the $B$ meson and lepton in units of $e$.
With the aim to render the transfer of our analysis to the positively charged channel more transparent, we will explicitly retain factors of $Q_B = Q_\ell$ in our formulae; it is, however, to be kept in mind that further modifications of the spinor structure apply beyond this simple alteration.
In \autoref{eq:matrix_element_B_to_l_nu_gamma}, we moreover abbreviate the leptonic matrix element $L_\nu = \bar{u}_\ell \gamma_\nu (1-\gamma_5) v_{\bar{\nu}}$ and introduce the hadronic tensor $T_\had^{\mu \nu}(k,q)$,
\beq\label{eq:definition_hadronic_tensor}
    Q_B T_\had^{\mu \nu}(k,q) 
    = \int\dfx \ex^{\iu q x} \braket{0 | \tior\{J_\elmag^\mu(x) J_\had^\nu(0)\} | B^-},
\eeq
and the FSR tensor $T_\FSR^\mu(p_\ell,p_\nu,q)$,
\begin{align}\label{eq:definition_FSR_tensor}
	Q_\ell T_\FSR^{\mu}(p_\ell,p_\nu,q) 
	= -\iu f_B p_\nu \int\dfx \ex^{\iu q x} \braket{\ell^- \bar{\nu}_\ell | \tior\{J_\elmag^\mu(x) J_\weak^\nu(0)\} | 0}.
\end{align}
While the hadronic tensor $T_\had^{\mu \nu}(k,q)$ describes the genuinely nonperturbative physics of the process, $T_\FSR^{\mu}(p_\ell,p_\nu,q)$ comprises the FSR from the charged lepton and can be reduced to the $B$-meson decay constant $f_B$ and an entirely perturbative remainder.
The former can be decomposed into a set of \LN{Lorentz} structures and associated scalar-valued functions, which are commonly referred to as the $B \to \gamma^*$ form factors.
The purpose of this work is to study these form factors within a dispersive framework, which requires knowledge of their singularity structure in the two independent kinematic variables and of the form factors' asymptotic behavior, see \autoref{sec:form_factors}.

For the FSR tensor in the case of a massless charged lepton, one finds the remarkably simple result~\cite{Khodjamirian:2001ga,Beneke:2011nf,Beneke:2021rjf,Ivanov:2021jsr,Janowski:2021yvz}
\beq\label{eq:FSR_tensor_massless}
    T_{\FSR,0}^\mu(p_\ell,p_\nu,q) 
    = f_B L^\mu.
\eeq
The case of nonzero mass leads to the more intricate formula~\cite{Bijnens:1992en,Bijnens:1994me}
\beq\label{eq:FSR_tensor_massive}
    T_{\FSR,m_\ell}^\mu(p_\ell,p_\nu,q) 
    = f_B \bigg[ L^\mu + m_\ell \bar{u}_\ell \frac{2p_\ell^\mu + \gamma^\mu \slashed{q}}{(p_\ell + q)^2 - m_\ell^2} (1-\gamma_5) v_{\bar{\nu}} \bigg].
\eeq
For our purpose, it proves convenient to bring the FSR contribution into such a form that it shares a common factor of $L_\nu$ with its hadronic counterpiece, \Lat{i.e.},
\beq\label{eq:matrix_element_B_to_l_nu_gamma_rewritten}
	\braket{\ell^- \bar{\nu}_\ell \gamma^* | J_\weak^\nu(0) {J_\had}_\nu(0) | B^-}
	= e Q_B \eps_\mu^* \big[ T_\had^{\mu \nu}(k,q) + T_\FSR^{\mu \nu}(p_\ell,p_\nu,q) \big] L_\nu.
\eeq
It is straightforward to achieve such a description for the massless case, $m_\ell = 0$, \autoref{eq:FSR_tensor_massless}.
For the massive case, $m_\ell \neq 0$, we make use of the \LN{Chisholm} identity~\cite{Pal:2007dc}
\beq
    \iu \eps^{\mu \nu \rho \sigma} \gamma_\sigma \gamma_5 
    = \gamma^\mu \gamma^\nu \gamma^\rho - g^{\mu \nu} \gamma^\rho + g^{\mu \rho} \gamma^\nu - g^{\nu \rho} \gamma^\mu,
\eeq
with the convention $\eps^{0123} = +1$.
From this, we obtain
\beq\label{eq:FSR_tensor_massive_rewritten}
    T_{\FSR}^{\mu \nu}(p_\ell,p_\nu,q) 
    = f_B \bigg[ g^{\mu \nu} + \frac{2p_\ell^\mu p_\ell^\nu + p_\ell^\mu q^\nu + q^\mu p_\ell^\nu - (p_\ell \cdot q) g^{\mu \nu} + \iu \eps^{\mu \nu \rho \sigma} (p_\ell)_\rho q_\sigma}{(p_\ell+q)^2 - m_\ell^2} \bigg],
\eeq
which is valid only when contracted with the leptonic matrix element $L_\nu$.\footnote{%
    Note that one can, in principle, further make the replacement $p_\ell^\nu \to k^\nu$ in \autoref{eq:FSR_tensor_massive_rewritten} by virtue of the \LN{Dirac} equation for the neutrino.}
    
Because of gauge invariance, the full amplitude complies with the \LN{Ward} identity
\beq
    q_\mu \big[ T_\had^{\mu \nu}(k,q) + T_\FSR^{\mu \nu}(p_\ell,p_\nu,q) \big] L_\nu 
    = 0.
\eeq
However, the hadronic and FSR tensor are not individually gauge invariant but satisfy~\cite{Ivanov:2021jsr,Beneke:2011nf,Beneke:2021rjf}
\begin{align}\label{eq:inhomogeneous_gauge_invariance}
    q_\mu T_\had^{\mu \nu}(k,q) 
    &= -f_B (k+q)^\nu, \notag \\
    q_\mu T_\FSR^{\mu \nu}(p_\ell,p_\nu,q) 
    &= f_B (k+q)^\nu,
\end{align}
so that gauge invariance only holds for the sum of both contributions.
Based on \autoref{eq:inhomogeneous_gauge_invariance}, we split the hadronic tensor into a homogeneous part and an inhomogeneous part by means of $T_\had^{\mu \nu}(k,q) = T_{\had,\hom}^{\mu \nu}(k,q) + T_{\had,\inhom}^{\mu \nu}(k,q)$, which obey
\begin{align}\label{eq:hadronic_tensor_split}
    q_\mu T_{\had,\hom}^{\mu \nu}(k,q) 
    &= 0, \notag \\
    q_\mu T_{\had,\inhom}^{\mu \nu}(k,q) 
    &= -f_B (k+q)^\nu.
\end{align}
We have not yet made any choice of \LN{Lorentz} decomposition for $T_{\had}^{\mu\nu}(k,q)$ or its (in)homogeneous part.
In App.~\ref{app:inhomogeneities}, we demonstrate that any choice for the decomposition of the hadronic tensor leads to the relation
\beq\label{eq:contracting_homogeneous_hadronic_tensor_with_k}
    k_\nu T_{\had,\hom}^{\mu \nu}(k,q) 
    = T_\ps^\mu(k,q) + f_B (k+q)^\mu - k_\nu T_{\had,\inhom}^{\mu \nu}(k,q),
\eeq
where the pseudoscalar tensor $T_\ps^\mu(k,q)$ is defined in terms of the pseudoscalar weak current $J_P(x) = \bar{u}(x) \gamma_5 b(x)$ via
\beq\label{eq:definition_pseudoscalar_tensor}
    Q_B T_\ps^\mu(k,q) 
    = (m_b + m_u) \int\dfx \ex^{\iu q x} \braket{0 | \tior\{J_\elmag^\mu(x) J_\ps(0)\} | B^-},
\eeq
with $m_b$ and $m_u$ the $\overline{\text{MS}}$ masses of the $b$- and $u$-quarks.
As also shown in App.~\ref{app:inhomogeneities}, this tensor is not gauge invariant but, similar to \autoref{eq:inhomogeneous_gauge_invariance}, fulfills
\beq\label{eq:contracting_pseudoscalar_tensor_with_q}
    q_\mu T_\ps^\mu(k,q) 
    = -f_B \mB^2.
\eeq
For this reason, we proceed in analogy to \autoref{eq:hadronic_tensor_split} and split $T_\ps^\mu(k,q) = T_{\ps,\hom}^\mu(k,q) + T_{\ps,\inhom}^\mu(k,q)$, where
\begin{align}\label{eq:pseudoscalar_tensor_split}
    q_\mu T_{\ps,\hom}^\mu(k,q) 
    &= 0, \notag \\
    q_\mu T_{\ps,\inhom}^\mu(k,q) 
    &= -f_B \mB^2.
\end{align}
In this work, we additionally impose that the homogeneous part of the hadronic tensor fulfills
\beq\label{eq:condition_pseudoscalar_homogeneous_part}
    k_\nu T_{\had,\hom}^{\mu \nu}(k,q) 
    \overset{!}{=} T_{\ps,\hom}^\mu(k,q),
\eeq
which, using \autoref{eq:contracting_homogeneous_hadronic_tensor_with_k}, leads to the condition
\beq\label{eq:condition_pseudoscalar_inhomogeneous_part}
    T_{\ps,\inhom}^\mu(k,q) + f_B (k + q)^\mu - k_\nu T_{\had,\inhom}^{\mu \nu}(k,q) 
    = 0.
\eeq
This choice is natural because it relates one of the hadronic form factors of the axial-vector current with that of the pseudoscalar current, as is the case for hadronic form factors in other weak transitions, too.

The tensors $T_{\had}^{\mu\nu}(k,q)$ and $T_{\FSR}^{\mu\nu}(p_\ell,p_\nu,q)$ emerge in predictions for the decay $B^-(p) \to \ell^-(p_\ell) \bar{\nu}_\ell(p_\nu) \ellp(q_1) \ellpbar(q_2)$, with $\ell' \neq \ell$, $q = q_1 + q_2$,
\begin{align}\label{eq:M_B_to_3l_nu}
	\M(B^- \to \ell^- \bar{\nu}_\ell \ellp \ellpbar)
	&= \frac{4 \FC V_{ub}}{\sqrt{2}} \braket{\ell^- \bar{\nu}_\ell \ellp \ellpbar | \Op{V,L} | B^-} \notag \\
	&= \frac{\FC V_{ub}}{\sqrt{2}} \frac{e^2}{q^2} Q_B \big[ T_\had^{\mu \nu}(k,q) + T_\FSR^{\mu \nu}(p_\ell,p_\nu,q) \big] l_\mu L_\nu,
\end{align}
where we abbreviate the leptonic matrix element $l_\mu = \bar{u}_{\ell'} \gamma_\mu v_{\bar{\ell'}}$.
The discussion of the decay with identical lepton flavors, $\ell' = \ell$, is more involved~\cite{Beneke:2021rjf,Ivanov:2022uum}, since an additional diagram has to be taken into account due to the interchangeability of two final-state fermions, which is beyond the scope of this article.

\section{\boldmath{$B \to \gamma^*$} form factors}
\label{sec:form_factors}
We develop a method that closely resembles the BTT procedure~\cite{Tarrach:1975tu,Bardeen:1968ebo} to parameterize the homogeneous part of the hadronic tensor, see App.~\ref{app:BTT}.
Compared to the BTT procedure, our method has the advantage that the emerging form factors have definite angular-momentum and parity quantum numbers.
Our result reads
\begin{align}\label{eq:BTT_decomposition}
	T_{\had,\hom}^{\mu \nu}(k,q) 
	&= \frac{1}{\mB} [ (k \cdot q) g^{\mu \nu} - k^\mu q^\nu ] \F_1(k^2,q^2) + \frac{1}{\mB} \Big[ \frac{q^2}{k^2} k^\mu k^\nu - \frac{k \cdot q}{k^2} q^\mu k^\nu + q^\mu q^\nu - q^2 g^{\mu \nu} \Big] \F_2(k^2,q^2) \notag \\
	&\quad + \frac{1}{\mB} \Big[ \frac{k \cdot q}{k^2} q^\mu k^\nu - \frac{q^2}{k^2} k^\mu k^\nu \Big] \F_3(k^2,q^2) + \frac{\iu}{\mB} \eps^{\mu \nu \rho \sigma} k_\rho q_\sigma \F_4(k^2,q^2),
\end{align}
where the form factors $\F_1(k^2,q^2)$ and $\F_2(k^2,q^2)$ have axial-vector, $\F_3(k^2,q^2)$ has pseudoscalar, and $\F_4(k^2,q^2)$ vector quantum numbers with respect to the weak current.\footnote{%
    Note that for on-shell photons, only the form factors $\F_1(k^2,q^2)$ and $\F_4(k^2,q^2)$ contribute, which correspond to transverse polarizations.}
Assuming no modification due to the inhomogeneous part $T_{\had,\inhom}^{\mu \nu}(k,q)$, our form factors are free of kinematic singularities in $k^2$ and $q^2$ as well as kinematic zeroes in $q^2$.
However, to ensure a finite amplitude at $k^2 = 0$, the relation $\F_2(0,q^2) = \F_3(0,q^2)$ must hold for all $q^2$.
The factors of $\mB$ and the imaginary unit in \autoref{eq:BTT_decomposition} render the form factors dimensionless and---with the phase of the $B$ meson chosen appropriately---real-valued below the onset of the first branch cut.

The relations given in \autoref{eq:hadronic_tensor_split} constrain the inhomogeneous part of the hadronic tensor to the generic form
\beq\label{eq:generic_inhomogeneous_part}
    T_{\had,\inhom}^{\mu \nu}(k,q) 
    = - f_B \bigg[ a g^{\mu \nu} + b \frac{k^\mu k^\nu}{k \cdot q} + c \frac{k^\mu q^\nu}{k \cdot q} + (1-b) \frac{q^\mu k^\nu}{q^2} + (1-a-c) \frac{q^\mu q^\nu}{q^2} \bigg],
\eeq
where $a \equiv a(k^2,q^2)$, $b \equiv b(k^2,q^2)$, and $c \equiv c(k^2,q^2)$ are arbitrary real-valued coefficients.
The \LN{Levi}-\LN{Civita} tensor is absent in this expression because it carries the wrong quantum numbers in light of the fact that the inhomogeneity is entirely due to the axial-vector part of \autoref{eq:definition_hadronic_tensor}.
On account of \autoref{eq:pseudoscalar_tensor_split}, the inhomogeneous part of the pseudoscalar tensor furthermore takes the generic form
\beq\label{eq:generic_pseudoscalar_inhomogeneous_part}
    T_{\ps,\inhom}^\mu(k,q) 
    = -f_B \mB^2 \bigg[ d \frac{k^\mu}{k \cdot q} + (1 - d) \frac{q^\mu}{q^2} \bigg],
\eeq
where $d \equiv d(k^2,q^2)$ is an arbitrary real-valued coefficient.
Adopting the condition imposed in \autoref{eq:condition_pseudoscalar_inhomogeneous_part}, we find that
\beq\label{eq:fixed_pseudoscalar_inhomogeneous_part}
    d 
    = \frac{(1+a+c) (k \cdot q) + b k^2}{\mB^2},
\eeq
which fixes $T_{\ps,\inhom}^\mu(k,q)$ once $T_{\had,\inhom}^{\mu \nu}(k,q)$ is specified.
We collect four different choices for the coefficients, labeled $\LabelA$ through $\LabelD$, in~\autoref{tab:inhomogeneous_part_literature}.
With regard to the dispersive treatment of the form factors in this article, \Lat{i.e.}, the requirement of their singularity-free structure, the question emerges what an appropriate choice for these coefficients is.
\begin{table}[t]
	\begin{tabular}{c  c  c  c  c  c  c  c}
	\toprule
		Label & $a$ & $b$ & $c$ & $T_{\had,\inhom}^{\mu \nu}(k,q)$ & $d$ & $T_{\ps,\inhom}^\mu(k,q)$ & References \\
		\midrule
		$\LabelA$ & $1$ & $\frac{2(k \cdot q)}{2(k \cdot q) + q^2}$ & $0$ & $-f_B \Big[ g^{\mu \nu} + \frac{(2k^\mu +q^\mu) k^\nu}{2(k \cdot q) + q^2} \Big]$ & $\frac{2(k \cdot q)}{2(k \cdot q) + q^2}$ & $-f_B \mB^2 \frac{2k^\mu + q^\mu}{2(k \cdot q) + q^2}$ & \cite{Bardin:1976wv,Bijnens:1992en,Bijnens:1994me,Janowski:2021yvz}\\
		\\[-1em] 
		$\LabelB$ & $0$ & $\frac{k \cdot q}{k \cdot q + q^2}$ & $\frac{k \cdot q}{k \cdot q + q^2}$ & $-f_B \frac{(k+q)^\mu (k+q)^\nu}{k \cdot q + q^2}$ & $\frac{k \cdot q}{k \cdot q +q^2}$ & $-f_B \mB^2 \frac{k^\mu + q^\mu}{k \cdot q + q^2}$ & \cite{Khodjamirian:2001ga,Beneke:2011nf}\\
		\\[-1em] 
		$\LabelC$ & $0$ & $1$ & $1$ & $-f_B \frac{k^\mu (k+q)^\nu}{k \cdot q}$ & $\frac{2(k \cdot q) + k^2}{2(k \cdot q) + k^2 + q^2}$ & $-f_B \Big[ \mB^2 \frac{k^\mu}{k \cdot q} - \frac{q^2 k^\mu - (k \cdot q) q^\mu}{k \cdot q} \Big]$ & \cite{Beneke:2021rjf}\\
		\\[-1em]
		$\LabelD$ & $0$ & $0$ & $0$ & $-f_B \frac{q^\mu (k+q)^\nu}{q^2}$ & $\frac{k \cdot q}{2(k \cdot q) + k^2 + q^2}$ & $-f_B \Big[ \mB^2 \frac{q^\mu}{q^2} - \frac{(k \cdot q) q^\mu - q^2 k^\mu}{q^2} \Big]$ & \cite{Ivanov:2021jsr}\\
		\\[-1em]
	\bottomrule
	\end{tabular}
	\caption{The ansätze for the inhomogeneous part of the hadronic tensor used in the literature, expressed as in \autoref{eq:generic_inhomogeneous_part} for specific choices of the coefficients $a$, $b$, and $c$.
	Also shown are the resulting inhomogeneous parts of the pseudoscalar tensor, \autoref{eq:generic_pseudoscalar_inhomogeneous_part}, and its associated coefficient $d$, \autoref{eq:fixed_pseudoscalar_inhomogeneous_part}.
	The basis for the homogeneous part of the hadronic tensor differs from our choice, \autoref{eq:BTT_decomposition}, in some of the references.
	A thorough discussion of the various choices can be found in the main text.}
	\label{tab:inhomogeneous_part_literature}
\end{table}

\begin{figure}[t]
    \centering
    \includegraphics[scale=0.875]{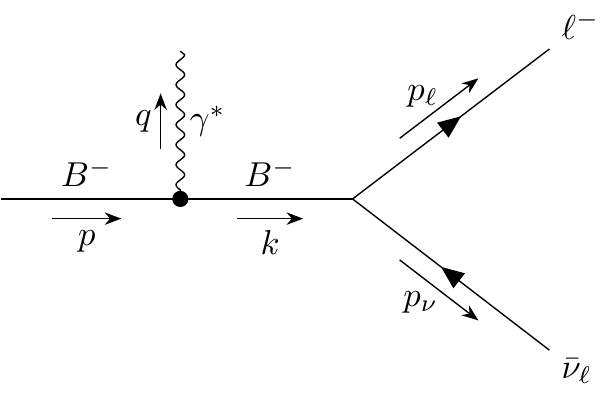}
    \caption{Diagram illustrating the $B$-meson pole in the variable $k^2$ as part of the hadronic tensor $T_\had^{\mu \nu}(k,q)$; see also the left diagram of \autoref{fig:B_to_l_nu_gamma_hadronic_diagrams}.}
    \label{fig:B_to_l_nu_gamma_hadronic_pole_diagram}
\end{figure}

Among the inhomogeneous parts of the hadronic tensor listed in~\autoref{tab:inhomogeneous_part_literature}, $\LabelA$ is the only choice that introduces a term singular in $[2(k \cdot q) + q^2] = (\mB^2 - k^2)$.
It is evident that this $k^2$-pole is associated with an intermediate $B$ meson~\cite{Bardin:1976wv}, as sketched in the left diagram of \autoref{fig:B_to_l_nu_gamma_hadronic_diagrams}, see also \autoref{fig:B_to_l_nu_gamma_hadronic_pole_diagram}.
The choices $\LabelB$ and $\LabelC$, on the other hand, introduce terms singular in $[(k \cdot q) + q^2]$ and $(k \cdot q)$, respectively, which correspond to $q^2$-dependent pole positions in the variable $k^2$; these are not associated with any hadronic intermediate state and are therefore not of dynamic but of kinematic origin.
Choice $\LabelD$ corresponds to a structure that is orthogonal to all BTT structures.
This might lead to the presumption that it leaves the form factors of \autoref{eq:BTT_decomposition} unaffected and thus free of kinematic singularities.
However, this choice exhibits a pole in $q^2$, which erroneously suggests the emergence of a dynamic photon pole; working at fixed order in quantum electrodynamics, such a pole cannot arise.
In fact, the behavior $\propto 1/q^2$ would lead to a double pole $\propto 1/q^4$ in \autoref{eq:M_B_to_3l_nu}, a feature that is to be avoided in any amplitude.
As a consequence of this double pole, choice $\LabelD$ is---in addition to the kinematic nature of the $q^2$ pole---disqualified by its effect on the longitudinal $B^- \to \ell^- \bar{\nu}_\ell \gamma^*$ helicity amplitude.

To further illustrate the effect that choice $\LabelD$ causes, we investigate the $B^- \to \ell^- \bar{\nu}_\ell \gamma^*$ amplitude in more detail.
From Eqs.~\eqref{eq:M_B_to_l_nu_gamma} and \eqref{eq:matrix_element_B_to_l_nu_gamma_rewritten}, one finds the squared spin-averaged amplitude for photons with polarization $\lambda$ to be given by
\begin{align}\label{eq:M_B_to_l_nu_gamma_squared}
    \lvert \overline{\M}(B^- \to \ell^- \bar{\nu}_\ell \gamma^*(\lambda)) \rvert^2 
    &= \frac{e^2 \FC^2 \lvert V_{ub} \rvert^2}{2} \eps_\mu^*(q; \lambda) \eps_\alpha(q; \lambda) \big[ T_\had^{\mu \nu}(k,q) + T_\FSR^{\mu \nu}(p_\ell,p_\nu,q) \big] \big[ T_\had^{\alpha \beta}(k,q) + T_\FSR^{\alpha \beta}(p_\ell,p_\nu,q) \big]^\dagger \notag \\
    &\quad \times \sum_{\text{spins}} L_\nu L_\beta^\dagger,
\end{align}
see App.~\ref{app:kinematics} for details on the kinematics.
For a longitudinal photon, $\lambda = 0$, this matrix element ought to vanish in the limit $q^2 \to 0$, \Lat{i.e.}, for an on-shell photon.
Using choice $\LabelD$, one does, however, find that the matrix element diverges $\propto f_B^2$, independent of any choice of form factors.
The discussion of such divergent contributions is not purely academic: in Ref.~\cite{Ivanov:2021jsr}, a supposed collinear enhancement of the $B^- \to \ell^- \bar{\nu}_\ell \ellp \ellpbar$ decay rate is discussed, which is caused by such an unphysical behavior as $q^2 \to 0$.
Therein, a different choice is made for the decomposition of the homogeneous tensor, in combination with choice $\LabelD$ for the inhomogeneous part and an inconsistent treatment of the charged lepton's finite mass in the FSR term.
Using the formulae of Ref.~\cite{Ivanov:2021jsr} and our result for the FSR tensor, \autoref{eq:FSR_tensor_massive_rewritten}, we validate that treating the effects of a finite lepton mass consistently resolves this issue and removes the supposed contribution due to a longitudinal on-shell photon.\footnote{%
    After submitting our article for review, this has been confirmed to us by the authors of Ref.~\cite{Ivanov:2021jsr} and is revised in an Erratum.}
This leads us to infer that the supposed collinear enhancement is not a physical feature of the $B^- \to \ell^- \bar{\nu}_\ell \ellp \ellpbar$ decay rate.

Moreover, we can draw conclusions from the results for the hadronic tensor in the decay $K^\pm \to \ell^\pm \nu_\ell \gamma^* (\to \ellp \ellpbar)$.
An explicit calculation in chiral perturbation theory at next-to-leading order~\cite{Bijnens:1992en,Bijnens:1994me} confirms that choice $\LabelA$ yields form factors that are free of kinematic singularities.
Transforming between choice $\LabelA$ and any other choice of \autoref{tab:inhomogeneous_part_literature} modifies the homogeneous part through introducing kinematic singularities.
Consequently, the assumption that choice $\LabelA$ leads to form factors free of kinematic singularities unavoidably implies the emergence of such singularities for all the other choices considered here.

Under some rather general, reasonable assumptions, it is possible to deduce that the inhomogeneous part of the hadronic tensor ought to be of the form
\beq
    T_{\had,\inhom}^{\mu \nu}(k,q) 
    = -f_B \bigg[ \hat{a} g^{\mu \nu} + \frac{(2k^\mu + q^\mu)k^\nu + (1-\hat{a})(2k^\mu + q^\mu)q^\nu}{2 (k \cdot q) + q^2} \bigg]
\eeq
in combination with the BTT basis of \autoref{eq:BTT_decomposition} for the homogeneous part.
Here, $\hat{a}$ is an arbitrary real-valued coefficient that does not depend on any of the momenta.
The assumptions underlying the above form are the following:
\begin{itemize}
    \item there exists a unique choice for the coefficients in \autoref{eq:generic_inhomogeneous_part} that leaves the form factors free of kinematic singularities;
    \item the apparent kinematic poles in $T_{\had,\inhom}^{\mu \nu}(k,q)$ cancel and no new such poles are introduced;
    \item a dynamic $B$-meson pole appears at most in the pseudoscalar form factor $\F_3(k^2,q^2)$.
\end{itemize}
Consequently, the inhomogeneous part of the pseudoscalar tensor, \autoref{eq:generic_pseudoscalar_inhomogeneous_part}, turns out to be given by
\beq
     T_{\ps,\inhom}^\mu(k,q) 
     = -f_B \bigg[\mB^2 \frac{2k^\mu + q^\mu}{2(k \cdot q) + q^2} - (1-\hat{a}) \frac{q^2 k^\mu - (k \cdot q) q^\mu}{2(k \cdot q) + q^2} \bigg].
\eeq
Assuming that $\hat{a}=1$ meets the above requirements, it can be shown that any other choice of $\hat{a}$ would introduce a dynamic pseudoscalar $B$-meson pole in the axial-vector form factors $\F_1(k^2,q^2)$ and $\F_2(k^2,q^2)$.
Since $\hat{a} = 1$ corresponds to the choice $\LabelA$ from \autoref{tab:inhomogeneous_part_literature}, this gives further indication that $\LabelA$ is the proper choice for our analysis.

For the reasons stated above, we make $\LabelA$ the default choice in the following and parameterize the hadronic tensor as
\beq\label{eq:full_hadronic_tensor}
	T_\had^{\mu \nu}(k,q) 
	= T_{\had,\hom}^{\mu \nu}(k,q) - f_B \bigg[ g^{\mu \nu} + \frac{(2k^\mu + q^\mu) k^\nu}{2(k \cdot q) + q^2} \bigg].
\eeq
This yields a total of six independent \LN{Lorentz} structures, which form a basis, see the discussion in the appendix of Ref.~\cite{Beneke:2021rjf}.
Having such a basis of structures allows us to find projectors $\proj_i^{\mu \nu}(k,q)$ that fulfill 
\beq
    {\proj_i}_{\mu \nu}(k,q) T_\had^{\mu \nu}(k,q) 
    =
    \begin{cases}
        \F_i(k^2,q^2), & i=1,\ldots,4, \\
        f_B/\mB, & i=5,6.
    \end{cases}
\eeq
Explicit formulae for these projectors are provided in App.~\ref{app:projectors}.

\section{Dispersion relations and \boldmath{$z$} expansion}
\label{sec:dispersion_relations}
We aim to parameterize the form factors $\F_i(k^2,q^2)$, $i=1,\ldots,4$, in accordance with analyticity and unitarity.
To this end, we split the form factors with respect to the photon's isospin according to $\F_i(k^2,q^2) = \F_i^{I=0}(k^2,q^2) + \F_i^{I=1}(k^2,q^2)$.
For each component, we then establish a set of dispersion relations and assume the underlying discontinuities to be dominated by the one-body intermediate states $\omega$ and $\rho$, respectively, which allows us to relate the $B \to \gamma^*$ form factors to the $B \to V$, $V=\omega, \rho$, analogs.
In doing so, we neglect contributions due to $B \to \phi$ in the isoscalar components for two reasons: first, these contributions are expected to be small due to the \LN{Okubo}--\LN{Zweig}--\LN{Iizuka} mechanism~\cite{Okubo:1963fa,Zweig:1964jf,Iizuka:1966fk}, and second, we lack nonperturbative input for the $B \to \phi$ form factors.
We also do not model contributions from further excited states, such as $\omega(1420)$ and $\rho(1450)$.
As a consequence, we provide our nominal phenomenological results only in the region $q^2 \lesssim 1\GeV^2$.

Based on \autoref{eq:definition_hadronic_tensor}, the discontinuity of the form factors with respect to $q^2$ and for fixed $k^2$ is given by~\cite{Colangelo:2000dp,Khodjamirian:2020btr}
\begin{align}
    \discq[ Q_B \F_i(k^2,q^2)]
    &= \discq [{\proj_i}_{\mu \nu}(k,q) Q_B T_\had^{\mu \nu}(k,q)] \notag \\
    &= {\proj_i}_{\mu \nu}(k,q) \Big[\iu \sum_n \int \dff \tau_n \, (2\pi)^4 \delta^{(4)}(q - P_n) \braket{0 | J_\elmag^\mu(0) | n} \braket{n | J_\had^\nu(0) | B^-} \Big].
\end{align}
Here, we use the $n$-body phase-space volume
\beq
    \dff \tau_n 
    = \prod_j \frac{\dff^3 p_j}{(2\pi)^3 2p_j^0} 
    = \prod_j \frac{\dff^4 p_j}{(2\pi)^4} \, (2\pi) \delta( p_j^2 - M_j^2 ) \theta(p_j^0)
\eeq
and $P_n = \sum_j p_j$ is the total momentum of the intermediate state.
Assuming the discontinuities of the isoscalar and isovector components to be dominated by the one-body intermediate states $\omega$ and $\rho$, respectively, we use
\beq
    \int \dff \tau_n \, (2\pi)^4 \delta^{(4)}(q - P_n) f(P_n) 
    = 2\pi \delta( q^2 - M_n^2 ) f(q)
\eeq
for the one-body phase-space volume to obtain
\beq\label{eq:discontinuities_isoscalar_isovector}
	\discq[ Q_B \F_i^I(k^2,q^2)]
	= {\proj_i}_{\mu \nu}(k,q) \Big[ 2\pi \iu \sum_{\lambda} \delta( q^2 - M_V^2 ) \braket{0 | J_\elmag^\mu(0) | V(q,\lambda)} \braket{V(q,\lambda) | J_\had^\nu(0) | B^-} \Big],
\eeq
with $V = \omega$ for $I=0$ and $V = \rho$ for $I=1$.
For the above matrix elements, we employ~\cite{Bharucha:2015bzk}
\begin{align}\label{eq:V_matrix_elements}
	\braket{0 | J_\elmag^\mu(0) | V(q,\lambda)} 
	&= \frac{\eta^\mu}{c_V} d_V M_V f_V, \\
	\braket{V(q,\lambda) | J_\had^\nu(0) | B^-} 
	&= \frac{\eta_\alpha^*}{c_V} \big[ P_1^{\nu \alpha}(k,q) V^{B \to V}(k^2) + P_2^{\nu \alpha}(k,q) A_1^{B \to V}(k^2) + P_3^{\nu \alpha}(k,q) A_3^{B \to V}(k^2) + P_P^{\nu \alpha}(k,q) A_0^{B \to V}(k^2) \big], \notag
\end{align}
where the form factors $V^{B \to V}(k^2)$, $A_1^{B \to V}(k^2)$, $A_3^{B \to V}(k^2)$, and $A_0^{B \to V}(k^2)$ are given in the so-called traditional basis and account for a vector-, two axial-vector-, and a pseudoscalar-like $B \to V$ transition.
Furthermore, $d_\omega = Q_u + Q_d = 1/3$, $d_\rho = Q_u - Q_d = 1$, and the composition of the $\omega$ and $\rho$ wave function is accounted for by the factors $c_\omega = c_\rho = \sqrt{2}$.
The decay constant of the respective vector meson is denoted by $f_V$, and $\eta^\mu \equiv \eta^\mu(q; \lambda)$ represents the polarization vector of the incoming vector meson with momentum $q$ and polarization $\lambda$.
The structures in \autoref{eq:V_matrix_elements} are given by~\cite{Bharucha:2015bzk}
\begin{align}
	P_1^{\nu \alpha} 
	&= \frac{2 \iu}{\mB + M_V} \eps^{\nu \alpha \beta \gamma} q_\beta k_\gamma, \notag
	&
	P_2^{\nu \alpha} 
	&= -\frac{1}{\mB - M_V} \big[ ( \mB^2 - M_V^2 ) g^{\nu \alpha} - ( k^\nu + 2q^\nu ) k^\alpha \big], \notag
	\\
	P_3^{\nu \alpha} 
	&= \frac{2 M_V}{k^2} \bigg[ k^\nu - \frac{k^2}{\mB^2 - M_V^2} ( k^\nu + 2q^\nu ) \bigg] k^\alpha,
	&
	P_P^{\nu \alpha} 
	&= -\frac{2 M_V}{k^2} k^\nu k^\alpha,
\end{align}
where we adjusted the phases to our convention.
Using the additional relation~\cite{Horgan:2013hoa,Bharucha:2015bzk}
\beq
	A_{12}^{B \to V}(k^2)
	= \frac{k^2 (\mB + M_V) ( \mB^2 - k^2 + 3 M_V^2 ) A_1^{B \to V}(k^2) + 2 M_V \lambda_V(k^2) A_3^{B \to V}(k^2)}{16 \mB M_V^2 (\mB + M_V) (\mB - M_V)},
\eeq
where $\lambda_V(k^2) \equiv \lambda(\mB^2,k^2,M_V^2)$, with $\lambda(x,y,z) = x^2 + y^2 + z^2 - 2(xy+xz+yz)$ the \LN{Källén} function, we can express all form factors of \autoref{eq:V_matrix_elements} in terms of $V^{B \to V}(k^2)$, $A_1^{B \to V}(k^2)$, $A_{12}^{B \to V}(k^2)$, and $A_0^{B \to V}(k^2)$, which fulfill the exact relation~\cite{Bharucha:2015bzk}
\beq\label{eq:BSZ_form_factors_relation}
    A_0(0) 
    = \frac{8 \mB M_V A_{12}(0)}{\mB^2 - M_V^2}.
\eeq
The generic parameterization of $F^{B \to V}(k^2) \in \{V(k^2), A_1(k^2), A_{12}(k^2), A_0(k^2)\}$ in terms of a series expansion in the conformal variable
\beq
    z_V(t) 
    = \frac{\sqrt{t_+ - t} - \sqrt{t_+ - t_0}}{\sqrt{t_+ - t} + \sqrt{t_+ - t_0}}\bigg\rvert_{V=\omega,\rho},
\eeq
with $t_0 = (1 - \sqrt{1 - t_-/t_+}) t_+$ and $t_\pm = (\mB \pm M_V)^2$, is given by~\cite{Bharucha:2015bzk}
\beq\label{eq:z_expansion_BSZ}
    F^{B \to V}(k^2) 
    = R_{J^P}(k^2) \sum_{j \geq 0} \alpha_j^{F,V} [ z_V(k^2) - z_V(0) ]^j,
\eeq
where the series is truncated after three summands; this truncation is imposed on us by the $B \to V$ parameters provided in Ref.~\cite{Bharucha:2015bzk}.
Here, the expansion takes into account the dominant subthreshold poles of the $B \to V$ form factors through the term $R_{J^P} = (1-k^2/m_{J^P}^2)^{-1}$, where $J^P$ refers to the angular-momentum and parity quantum number of the respective form factor, see \autoref{tab:z_expansion_BSZ}.
\begin{table}[t]
	\centering
	\begin{tabular}{l  l  l  c  c  c  c  c  c}
	\toprule
		$F^{B \to V}(k^2)$ & $J^P$ & $m_{J^P}$ & $\alpha_0^{F,\omega}$ & $\alpha_1^{F,\omega}$ & $\alpha_2^{F,\omega}$ & $\alpha_0^{F,\rho}$ & $\alpha_1^{F,\rho}$ & $\alpha_2^{F,\rho}$ \\
		\midrule
		$V^{B \to V}(k^2)$ & $1^-$ & $\mBStar$ & $0.304(38)$ & $-0.83(29)$ & $1.7(1.2)$ & $0.327(31)$ & $-0.86(18)$ & $1.80(97)$ \\
		\\[-1em] 
		$A_1^{B \to V}(k^2)$ & $1^+$ & $\mBOne$ & $0.243(31)$ & $0.34(24)$ & $0.09(57)$ & $0.262(26)$ & $0.39(14)$ & $0.16(41)$ \\
		\\[-1em]
		$A_{12}^{B \to V}(k^2)$ & $1^+$ & $\mBOne$ & $0.270(40)$ & $0.66(26)$ & $0.28(98)$ & $0.297(35)$ & $0.76(20)$ & $0.46(76)$ \\
		\\[-1em] 
		$A_0^{B \to V}(k^2)$ & $0^-$ & $\mB$ & $0.328(48)$ & $-0.83(30)$ & $1.4(1.2)$ & $0.356(42)$ & $-0.83(20)$ & $1.3(1.0)$ \\
	\bottomrule
	\end{tabular}
	\caption{The quantum numbers $J^P$, resonance masses $m_{J^P}$, and numerical values (rounded to two significant digits) of the series coefficients $\alpha_j^{F,V}$~\cite{Bharucha:2015bzk} for the $z$ expansion of the form factors $F^{B \to V}(k^2)$, truncated after three summands, see \autoref{eq:z_expansion_BSZ}.
	The corresponding values of the resonance masses can be found in App.~\ref{app:constants}.
	Because of parity conservation of the strong interactions, no form factor with $J^P=0^+$ exists.
	For the exact numerical values of $\alpha_j^{F,V}$ and the covariances as well as correlations between these, see Ref.~\cite{Bharucha:2015bzk}.
	Note that $\alpha_0^{A_0,V}$ and $\alpha_0^{A_{12},V}$ are not independent but have to fulfill the exact relation given in \autoref{eq:BSZ_form_factors_relation}.}
	\label{tab:z_expansion_BSZ}
\end{table}

The isoscalar and isovector form factors can then be reconstructed from
\beq\label{eq:cauchys_integral_formula}
    Q_B \F_i^I(k^2,q^2) 
    = \frac{1}{2 \pi \iu} \int_{s_\text{thr}}^\infty \dff s \, \frac{\discs[ Q_B \F^I_i(k^2,s)]}{s - q^2},
\eeq
where $s_\text{thr} = 9 \Mpi^2, 4 \Mpi^2$ for $I = 0, 1$, respectively.
In the above, no subtractions are needed for convergence, since the discontinuities drop off as $1/q^2$ asymptotically; see App.~\ref{app:asymptotics}.
Inserting \autoref{eq:discontinuities_isoscalar_isovector} into \autoref{eq:cauchys_integral_formula} and using the polarization sum of the $\omega$ and $\rho$ mesons,
\beq
	\sum_{\lambda} \eta_\mu(q; \lambda) \eta_\nu^*(q; \lambda) 
	= -g_{\mu \nu} + \frac{q_\mu q_\nu}{M_V^2},
\eeq
we obtain the VMD result for the $B \to \gamma^*$ form factors,
\begin{align}\label{eq:reconstructed_form_factors}
    Q_B \F_1^I(k^2,q^2) 
    &= \mB M_V f_V d_V \frac{16 \mB M_V^2 A_{12}^{B \to V}(k^2) - (\mB + M_V) ( \mB^2 - k^2 - M_V^2 ) A_1^{B \to V}(k^2)}{ \lambda_V(k^2) (q^2 - M_V^2)}, \notag \\
    Q_B \F_2^I(k^2,q^2) 
    &= 2 \mB M_V f_V d_V \frac{4 \mB ( \mB^2 - k^2 - M_V^2 ) A_{12}^{B \to V}(k^2) - (\mB + M_V) k^2 A_1^{B \to V}(k^2)}{ \lambda_V(k^2) (q^2 - M_V^2)}, \notag \\
    Q_B \F_3^I(k^2,q^2) 
    &= \mB f_V d_V \frac{A_0^{B \to V}}{q^2 - M_V^2}, \notag \\
    Q_B \F_4^I(k^2,q^2) 
    &= \mB M_V f_V d_V \frac{V^{B \to V}(k^2)}{(\mB + M_V) (q^2 - M_V^2)}.
\end{align}
Compared to $\F_1(k^2,q^2)$ and $\F_4(k^2,q^2)$, the form factors $\F_2(k^2,q^2)$ and $\F_3(k^2,q^2)$ enter observables with a relative suppression factor of $q^2$, thereby ensuring that unphysical longitudinal on-shell photons do not contribute.

\begin{figure}[t]
    \centering
    \includegraphics[scale=0.625]{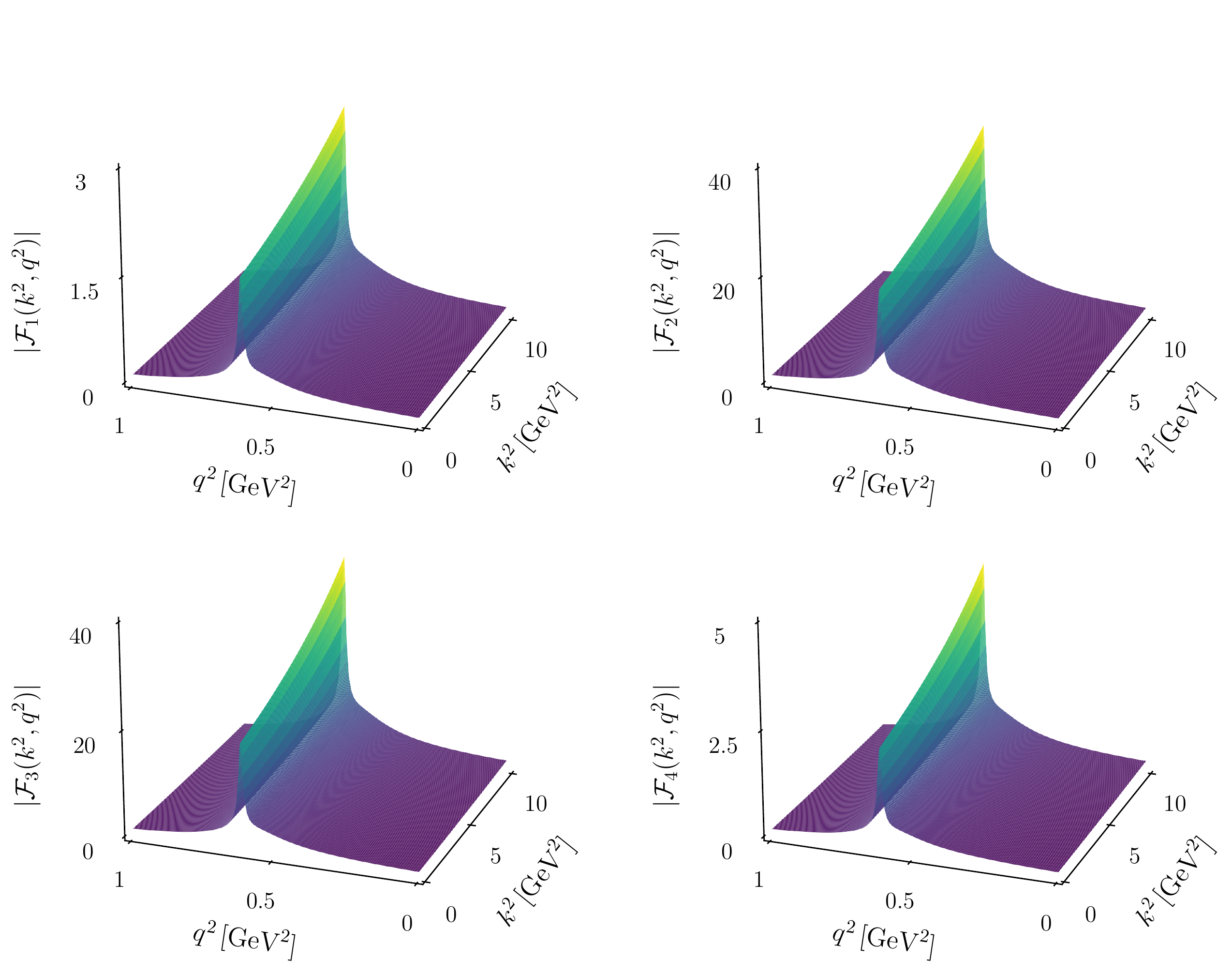}
    \caption{Three-dimensional plots showing the absolute values of the full form factors, \autoref{eq:full_B_to_gamma_form_factors}, in the range $k^2 \in [0,10]\GeV^2$ and $q^2 \in [0,1] \GeV^2$.
    The peak of the $\omega$ resonance is clearly visible, while the $\rho$ resonance is lower in magnitude and hardly discernible here.}
    \label{fig:plots_3D}
\end{figure}

\begin{figure}[t]
    \centering
    \includegraphics[scale=0.5]{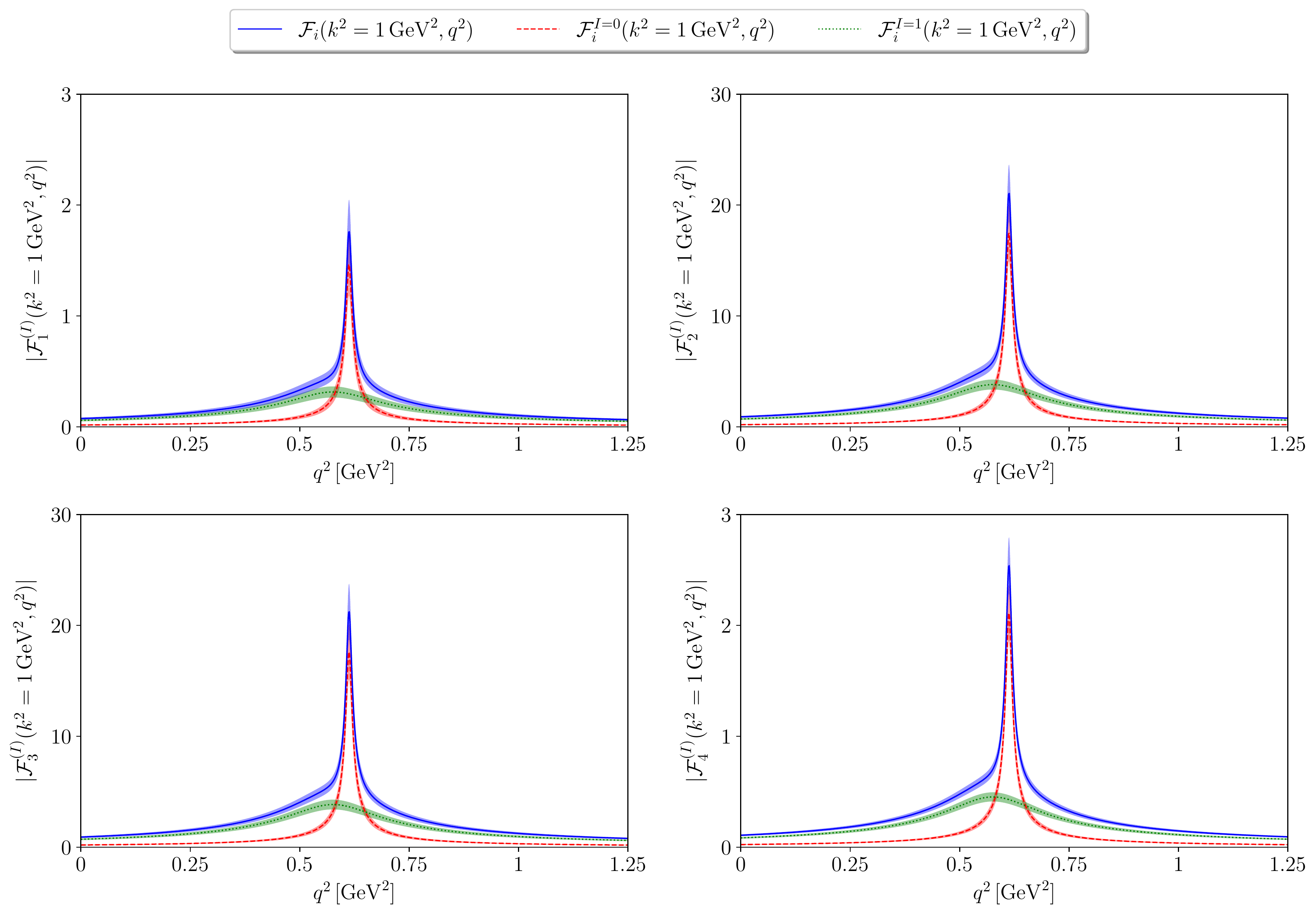}
    \caption{Two-dimensional plots of the absolute values of the form factors' isoscalar and isovector components as well as the sum of these for $k^2 = 1\GeV$ fixed in the range $q^2 \in [0,1.25] \GeV$.
    We additionally show the uncertainties of the corresponding contributions.}
    \label{fig:plots_2D}
\end{figure}

Naturally, we now aim to use an expansion similar to \autoref{eq:z_expansion_BSZ} for the $B \to \gamma^*$ form factors,
\beq\label{eq:z_expansion_form_factors}
   Q_B \F_i^I(k^2,q^2) 
   = R_{J^P}(k^2) \sum_{j \geq 0} \beta_{i,j}^V(q^2) [ z_V(k^2) - z_V(0) ]^j, 
\eeq
where the form factors have definite angular-momentum and parity assignments, with the term $R_{J^P}(k^2)$ again accounting for the dominant subthreshold poles in the variable $k^2$.
In contrast to \autoref{eq:z_expansion_BSZ}, the series coefficients have a dependence on $q^2$, for which we will assume VMD and use an \Lat{ad hoc} \LN{Breit}--\LN{Wigner} (BW) ansatz with the resonance's width inserted by hand,
\beq\label{eq:form_factors_series_coefficients}
    \beta_{i,j}^V(q^2) 
    = N_{i,j}^V P_V^\BW(q^2).
\eeq
At this, it is justified to use a monopole-like ansatz because the form factors drop off as $1/q^2$ asymptotically; see App.~\ref{app:asymptotics}.
Because of its smallness, we use a constant approximation for the $\omega$ decay width above the $3\pi$ threshold, whereas we incorporate the broad $\rho$ width energy-dependently,
\begin{align}
    P_\omega^\BW(q^2) 
    &= \frac{\Momega^2}{\Momega^2 - q^2 - \iu \Momega \, \Gamma_\omega},
    &
    P_\rho^\BW(q^2) 
    &= \frac{\Mrho^2}{\Mrho^2 - q^2 - \iu \sqrt{q^2} \, \Gamma_\rho(q^2)}.
\end{align}
Here, the proper threshold behavior is implied for the $\omega$, \Lat{i.e.}, $\Gamma_\omega=0$ for $q^2 < 9\Mpi^2$, and the energy-dependent width of the $\rho$ is parameterized according to~\cite{Zanke:2021wiq}
\begin{align}
    \Gamma_\rho(q^2) 
    &= \theta(q^2 - 4\Mpi^2) \frac{\gamma_{\rho \to \pi \pi}(q^2)}{\gamma_{\rho \to \pi \pi}(\Mrho^2)} \Gamma_\rho, 
    & 
    \gamma_{\rho \to \pi \pi}(q^2) 
    &= \frac{(q^2 - 4\Mpi^2)^{3/2}}{q^2}.
\end{align}
The normalizations $N_{i,j}^V$ can be determined from \autoref{eq:reconstructed_form_factors} by inserting Eqs.~\eqref{eq:z_expansion_BSZ} and \eqref{eq:z_expansion_form_factors} and using the numerical values from \autoref{tab:z_expansion_BSZ} to match at $q^2 = 0$, resulting in \autoref{tab:z_expansion_form_factors}.
The full form factors are then given by
\begin{align}\label{eq:full_B_to_gamma_form_factors}
    Q_B \F_i(k^2,q^2) 
    &= Q_B [\F_i^{I=0}(k^2,q^2) + \F_i^{I=1}(k^2,q^2)] \notag \\
    &= R_{J^P}(k^2) \sum_{\substack{V=\omega,\rho\\ j \geq 0}} N_{i,j}^V P_V^\BW(q^2) [ z_V(k^2) - z_V(0) ]^j.
\end{align}
We present three-dimensional plots of the absolute values of the full form factors, \autoref{eq:full_B_to_gamma_form_factors}, in \autoref{fig:plots_3D}.
In addition, we present two-dimensional plots in \autoref{fig:plots_2D}, where we also show the absolute values of the isoscalar and isovector components separately, \autoref{eq:z_expansion_form_factors}, including uncertainties and with $k^2 = 1 \GeV$ fixed.
\begin{table}[t]
	\centering
	\begin{tabular}{l  l  l  c  c  c  c  c  c}
	\toprule
		$\F_i(k^2,q^2)$ & $J^P$ & $m_{J^P}$ & $N_{i,0}^\omega$ & $N_{i,1}^\omega$ & $N_{i,2}^\omega$ & $N_{i,0}^\rho$ & $N_{i,1}^\rho$ & $N_{i,2}^\rho$ \\
		\midrule
		$\F_1(k^2,q^2)$ & $1^+$ & $\mBOne$ & $0.0156(30)$ & $-0.033(19)$ & $0.003(85)$ & $0.0557(88)$ & $-0.115(48)$ & $0.01(24)$ \\
		\\[-1em] 
		$\F_2(k^2,q^2)$ & $1^+$ & $\mBOne$ & $-0.186(27)$ & $0.39(14)$ & $-0.17(52)$ & $-0.676(79)$ & $1.34(41)$ & $-0.6(1.5)$ \\
		\\[-1em]
		$\F_3(k^2,q^2)$ & $0^-$ & $\mB$ & $-0.186(27)$ & $0.47(17)$ & $-0.80(71)$ & $-0.676(79)$ & $1.58(39)$ & $-2.5(2.0)$ \\
		\\[-1em]
		$\F_4(k^2,q^2)$ & $1^-$ & $\mBStar$ & $-0.0222(28)$ & $0.061(21)$ & $-0.125(91)$ & $-0.0795(75)$ & $0.209(44)$ & $-0.44(23)$ \\
	\bottomrule
	\end{tabular}
	\caption{The quantum numbers $J^P$, resonance masses $m_{J^P}$, and numerical values (rounded to two significant digits) of the normalizations $N_{i,j}^V$ for the $z$ expansion of the form factors $\F_i(k^2,q^2)$, truncated after three summands, see \autoref{eq:z_expansion_form_factors}.
	The corresponding values of the resonance masses can be found in App.~\ref{app:constants}.
	For the covariances between the normalizations, see App.~\ref{app:intermediate_results}.
	Note that $N_{2,0}^V$ and $N_{3,0}^V$ are identical due to the exact relation given in \autoref{eq:BSZ_form_factors_relation} or, equivalently, the condition $\F_2(0,q^2) = \F_3(0,q^2)$ imposed below \autoref{eq:BTT_decomposition}.}
	\label{tab:z_expansion_form_factors}
\end{table}

\section{Phenomenology}
\label{sec:pheno}
The decay $B^- \to \ell^- \bar{\nu}_\ell \ellp \ellpbar$ provides a rich phenomenology through a large number of angular observables.
They arise from the differential decay width $\dff \Gamma \equiv \dff \Gamma(B^- \to \ell^- \bar{\nu}_\ell \ellp \ellpbar)$, which is given by
\beq
    \dff\Gamma 
    = \frac{1}{2 \mB} \lvert \overline{\M} \rvert^2 \dff \Phi_4(p; p_\ell,p_\nu,q_1,q_2),
\eeq
where $\lvert \overline{\M} \rvert^2 \equiv \lvert \overline{\M}(B^- \to \ell^- \bar{\nu}_\ell \ellp \ellpbar) \rvert^2$ is the squared spin-average of \autoref{eq:M_B_to_3l_nu}.
The \LN{Lorentz}-invariant four-body phase space is conveniently split according to~\cite{ParticleDataGroup:2020ssz}
\beq\label{eq:four_body_phase_space}
    \dff \Phi_4(p; p_\ell,p_\nu,q_1,q_2) 
    = \dff \Phi_2(p;k,q) \dff \Phi_2(k;p_\ell,p_\nu) \dff \Phi_2(q;q_1,q_2) \frac{\dff k^2}{2\pi} \frac{\dff q^2}{2\pi}.
\eeq
Here, $\dff \Phi_2(p;k,q)$, $\dff \Phi_2(k;p_\ell,p_\nu)$, and $\dff \Phi_2(q;q_1,q_2)$ are the respective \LN{Lorentz}-invariant two-body phase space measures of the subsystems $\{\ell^- \bar{\nu}_\ell(k), \gamma^*(q)\}$, $\{\ell^-(p_\ell), \bar{\nu}_\ell(p_\nu)\}$, and $\{\ellp(q_1), \ellpbar(q_2)\}$.
The fivefold differential decay rate reads
\beq
    \frac{\dff^5 \Gamma}{\dff k^2 \dff q^2 \dff \cos\vartheta_W \dff \cos\vartheta_\gamma \dff \varphi} 
    = \frac{\lvert \vec{p}_\gamma \rvert \lvert \vec{p}_{\ell} \rvert \lvert \vec{p}_{\ell'} \rvert}{4096 \mB^2 \pi^6 \sqrt{k^2} \sqrt{q^2}} \lvert \overline{\M} \rvert^2,
\eeq
where $\vartheta_{W}$ is the polar angle of $\ell^-(p_\ell)$ in the center-of-mass system (CMS) $\{\ell^-(p_\ell), \bar{\nu}_\ell(p_\nu)\}$, $\vartheta_{\gamma}$ is the polar angle of $\ellp(q_1)$ in the CMS $\{\ellp(q_1), \ellpbar(q_2)\}$, and $\varphi$ is the relative azimuthal angle between the planes of these two subsystems.
Moreover, $\lvert \vec{p}_\gamma \rvert$, $\lvert \vec{p}_{\ell} \rvert$, and $\lvert \vec{p}_{\ell'} \rvert$ are the magnitudes of the three-momenta of the photon and the negatively charged leptons in the respective CMS; further details on the kinematics and the four-body phase space are provided in App.~\ref{app:kinematics}.
The angular integrations can be performed analytically, leading to
\begin{align}\label{eq:two_fold_decay_width}
    \frac{\dff^2 \Gamma}{\dff k^2 \dff q^2} 
    &= \mathcal{N} \bigg[ \sum_{i=1}^4 \frac{f_{i,i}}{\mB^2} \lvert \F_i(k^2,q^2) \rvert^2 + 2 \sum_{\substack{i=1\\j>i}}^4 \frac{f_{i,j}}{\mB^2} \Re[\F_i(k^2,q^2) \F_j^*(k^2,q^2)] + 2 f_B \sum_{i=1}^4 \frac{f_{i,5}}{\mB} \Re[\F_i(k^2,q^2)] + f_{5,5} f_B^2 \bigg], \notag \\
    \mathcal{N} 
    &= \frac{\FC^2 \lvert V_{ub} \rvert^2 e^4 \lvert \vec{p}_\gamma \rvert \lvert \vec{p}_{\ell} \rvert \lvert \vec{p}_{\ell'} \rvert}{8192 \mB^2 \pi^6 \sqrt{k^2} \sqrt{q^{10}}},
\end{align}
where an additional dependence of the lepton masses $m_{\ell^{(\prime)}}$ in the functions $f_{i,j} \equiv f_{i,j}(k^2,q^2)$ is omitted.
We collect the resulting expressions for these functions in App.~\ref{app:intermediate_results}.
The remaining integrations over $k^2$ and $q^2$ have to be performed numerically,
\beq\label{eq:decay_width}
    \Gamma 
    = \int \dff q^2 \int \dff k^2  \frac{\dff^2 \Gamma}{\dff k^2 \dff q^2},
\eeq
where the available phase space is bounded by $k^2 \in [m_\ell^2, (\mB - \sqrt{q^2})^2]$ and $q^2 \in [4m_{\ell'}^2, (\mB - m_\ell)^2]$.
Our results will be quoted for the branching ratio, $\BR = \Gamma \, \tau_B/\hbar$, where $\tau_B$ is the lifetime of the charged $B$ meson.

Beyond the integrated decay rate, another observable of interest is the FB asymmetry.
It provides a complementary probe of the form factors as compared to the decay width and is defined as
\beq
    A_\FB(k^2, q^2) 
    = \bigg(\frac{\dff^2 \Gamma}{\dff k^2 \dff q^2} \bigg)^{-1} \int\dff \cos\vartheta_W \, \sgn[\cos\vartheta_W] \frac{\dff^3 \Gamma}{\dff k^2 \dff q^2 \dff \cos\vartheta_{W}}.
\eeq
As for the decay width, the integration over the angle(s) can be performed analytically, with the result
\begin{align}\label{eq:FB_asymmetry}
    A_\FB(k^2, q^2) 
    &= \bigg(\frac{\dff^2 \Gamma }{\dff k^2 \dff q^2}\bigg)^{-1} \\
    &\quad \times \mathcal{N} \bigg[ \sum_{i=1}^4 \frac{g_{i,i}}{\mB^2} \lvert \F_i(k^2,q^2) \rvert^2 + 2 \sum_{\substack{i=1\\j>i}}^4 \frac{g_{i,j}}{\mB^2} \Re[\F_i(k^2,q^2) \F_j^*(k^2,q^2)] + 2 f_B \sum_{i=1}^4 \frac{g_{i,5}}{\mB} \Re[\F_i(k^2,q^2)] + g_{5,5} f_B^2 \bigg], \notag
\end{align}
where the functions $g_{i,j} \equiv g_{i,j}(k^2,q^2)$ also depend on the lepton masses $m_{\ell^{(\prime)}}$.
The resulting expressions for these functions are collected in App.~\ref{app:intermediate_results}.
Experimentally, it is convenient to access the integrated asymmetry, which is defined as
\beq\label{eq:integrated_FB_asymmetry}
    \langle A_\FB(k^2,q^2) \rangle
    = \bigg\langle \frac{\dff^2 \Gamma}{\dff k^2 \dff q^2} \bigg\rangle^{-1} \int\dff \cos\vartheta_W \, \sgn[\cos\vartheta_W] \bigg\langle \frac{\dff^3 \Gamma}{\dff k^2 \dff q^2 \dff \cos\vartheta_{W}} \bigg\rangle,
\eeq
where $\langle \cdots \rangle$ denotes the integration over a suitable bin in the kinematic variables $k^2$ and $q^2$.

We provide numerical results for both observables for the processes $B^- \to \ell^- \bar{\nu}_\ell \ellp \ellpbar$ with $\ell \in \{e,\mu,\tau\}$ and $\ell' \in \{e,\mu\}$ in \autoref{tab:numerical_results}.
Decays involving a $\tau^-\tau^+$ pair are not considered here, since the ditau threshold is large compared to our self-imposed upper cutoff in the variable $q^2$.
We do not provide results for the decay with $\ell' = \ell$ either, see the discussion at the end of \autoref{sec:hadronic_tensor}.
Our results are obtained 
\begin{enumerate}[label=(\roman*)]
    \item after integrating over the full phase space in $k^2$ and $q^2$;
    \item after integrating over the phase space with an upper cutoff at $q^2 = 1 \GeV^2$.
\end{enumerate}
Beyond the $q^2$ cutoff, the absence of the modeling of the $\phi$ and further resonances introduces a hardly quantifiable model uncertainty.
The latter variant therefore provides our nominal results.
Modeling the contributions beyond the cutoff seems possible in light of similar efforts in the case of $B \to \pi\pi$ form factors~\cite{Daub:2015xja,Ropertz:2018stk} and is left for future work.
\begin{table}[t]
	\centering
	\begin{tabular}{l  l  c  c}
	\toprule
		Process & Upper cutoff $q^2$ & $\BR$ & $A_\FB$ \\
		\midrule
		\multirow{2}{*}{$B^- \to e^- \bar{\nu}_e \mu^- \mu^+$} & None & $3.19(43)_N(25)_{V_{ub}} \times 10^{-8}$ & $-0.358(31)_N$ \\
		\\[-1em] 
		& $1 \GeV^2$ & $3.13(42)_N(25)_{V_{ub}} \times 10^{-8}$ & $-0.361(32)_N$ \\
		\midrule
		\multirow{2}{*}{$B^- \to \mu^- \bar{\nu}_\mu e^- e^+$} & None & $3.78(47)_N(30)_{V_{ub}} \times 10^{-8}$ & $-0.398(38)_N$ \\
		\\[-1em] 
		& $1 \GeV^2$ & $3.72(46)_N(30)_{V_{ub}} \times 10^{-8}$ & $-0.401(38)_N$ \\
        \midrule
        \multirow{2}{*}{$B^- \to \tau^- \bar{\nu}_\tau e^- e^+$} & None & $2.75(27)_N(22)_{V_{ub}} \times 10^{-8}$ & $-0.500(18)_N$ \\
		\\[-1em] 
		& $1 \GeV^2$ & $2.72(27)_N(22)_{V_{ub}} \times 10^{-8}$ & $-0.502(18)_N$ \\
        \midrule
        \multirow{2}{*}{$B^- \to \tau^- \bar{\nu}_\tau \mu^- \mu^+$} & None & $1.77(23)_N(14)_{V_{ub}} \times 10^{-8}$ & $-0.458(15)_N$ \\
		\\[-1em] 
		& $1 \GeV^2$ & $1.75(23)_N(14)_{V_{ub}} \times 10^{-8}$ & $-0.460(15)_N$ \\
	\bottomrule
	\end{tabular}
	\caption[]{%
    Numerical results for the branching ratio and FB asymmetry, see Eqs.~\eqref{eq:decay_width} and \eqref{eq:integrated_FB_asymmetry}, for $B^- \to \ell^- \bar{\nu}_\ell \ellp \ellpbar$ in the SM.
	The quoted uncertainties originate from the parametric uncertainties on the normalizations $N_{i,j}^V$ and $V_{ub}$, respectively.
	Because of the absence of $\CP$ violation in the SM, the results for the $\CP$-conjugated decay modes are identical.
	Within uncertainties, our predictions for the branching ratio of the process $B^- \to e^- \bar{\nu}_e \mu^- \mu^+$ agree well with Ref.~\cite{Ivanov:2021jsr}, $\BR(B^- \to e^- \bar{\nu}_e \mu^- \mu^+) = \{3.01 \times 10^{-8}, 2.96 \times 10^{-8}\}$, without and with an upper cutoff, respectively.
	For the process $B^- \to \mu^- \bar{\nu}_{\mu} e^- e^+$, however, our results are in strong tension with Ref.~\cite{Ivanov:2021jsr}, $\BR(B^- \to \mu^- \bar{\nu}_{\mu} e^- e^+) = \{6.38 \times 10^{-7}, 6.37 \times 10^{-7}\}$, which can be attributed to the unphysical collinear enhancement inferred therein,\footnote{%
        The tension with our result for the electron channel is reduced but not removed entirely with the results quoted in the Erratum to Ref.~\cite{Ivanov:2021jsr}.}
    see the discussion in \autoref{sec:form_factors}.
	The results of Ref.~\cite{Beneke:2021rjf}, Table 2, are---within their uncertainties---compatible with our results; note the numerically insignificant impact of the slight difference in the upper integration boundary used therein.
    }
    \label{tab:numerical_results}
\end{table}

\section{Summary and outlook}
\label{sec:summary_outlook}
In this article, we use dispersive methods to study the $B \to \gamma^*$ form factors underlying the decay $B^- \to \ell^- \bar{\nu}_\ell \ellp \ellpbar$, where we limit our analysis to the case $\ell' \neq \ell$.
We separate the full $B^- \to \ell^- \bar{\nu}_\ell \ellp \ellpbar$ amplitude into a nonperturbative hadronic tensor and a perturbative FSR piece and, in doing so, thoroughly investigate the properties of these individual objects.
One of the major advances of our analysis is to treat nonzero lepton masses consistently in the FSR piece at all stages.
The separation of the full amplitude into a hadronic tensor and an FSR piece leads to an ambiguity with regard to the dispersive treatment.
More specifically, it hinders one to find a decomposition into \LN{Lorentz} structures and form factors that are free of kinematic singularities.
As a remedy, we discuss in great detail how the hadronic tensor can be split into a homogeneous and an inhomogeneous part, with the homogeneous part being chosen such that it contains form factors with well-separated angular-momentum and parity quantum numbers.
From this, we propose a decomposition of the homogeneous part of the hadronic tensor into a set of \LN{Lorentz} structures and four form factors that are free of kinematic singularities in both the weak momentum and the photon momentum.
This renders possible a dispersive treatment of the form factors.
For the parameterization of the inhomogeneous part, we consider several choices from the literature and investigate their effect on the full amplitude in great detail, in particular with regard to the singularity-free property of the form factors.
Moreover, we find that the inhomogeneous part needs to be of a specific form under a few reasonable assumptions.
These considerations allow us to eliminate all except for one choice for the inhomogeneous part from the literature, which we consequently fix for the remainder of our analysis.

Having found a decomposition of the hadronic tensor into four form factors that are free of kinematic singularities, we split the form factors into their isospin components and establish a set of dispersion relations that relate the $B \to \gamma^*$ form factors to the well-known $B \to V$, $V = \omega, \rho$, analogs.
The $B \to V$ form factors are expanded in a series in the conformal variable $z(t)$, with the dominant subthreshold poles taken into account via a pole factor.
Performing a similar series expansion for the $B \to \gamma^*$ form factors and using a VMD ansatz for the virtual photon, we are able to parameterize these form factors reliably below the onset of the $\phi$.

Using our framework, we perform a phenomenological analysis by means of two observables: the branching ratio and the FB asymmetry.
The numerical results for these quantities are given for $\ell \neq \ell'$ and agree with previous determinations from the literature.

Possible future improvements of our framework involve the inclusion of the $\phi$ contribution and replacing the resonant $\rho$ by a description of the two-pion intermediate state, in which the $\rho$ can be included model-independently through pion--pion rescattering~\cite{Kang:2013jaa}.
The $B\to\gamma^*$ form factors are then obtained via a dispersion relation in a similar way to the reconstruction of, \Lat{e.g.}, the $\eta^{(\prime)}$ transition form factors from $\pi \pi$ intermediate states~\cite{Hanhart:2013vba, Holz:2022hwz}.


\begin{acknowledgments}
We are grateful to Yaroslav Kulii for helping with the translation of Ref.~\cite{Bardin:1976wv} from Russian to English.
We further thank Martin Beneke, Philipp B\"oer, Philip L\"ughausen, M\'eril Reboud, and K.~Keri Vos for useful discussions.
Financial support by the DFG through the funds provided to the Sino--German Collaborative Research Center TRR110 ``Symmetries and the Emergence of Structure in QCD'' (DFG Project-ID 196253076 -- TRR 110) is gratefully acknowledged.
The work of SK and DvD was further supported by the DFG within the Emmy Noether Programme under grant DY-130/1-1.
DvD acknowledges ongoing support by the UK Science and Technology Facilities Council (grant numbers ST/V003941/1 and ST/X003167/1).
\end{acknowledgments}


\appendix

\section{Inhomogeneous tensor identities}
\label{app:inhomogeneities}
In this appendix, we derive the identities for the hadronic tensor $T_\had^{\mu \nu}(k,q)$ and pseudoscalar tensor $T_\ps^{\mu}(k,q)$ given in Eqs.~\eqref{eq:contracting_homogeneous_hadronic_tensor_with_k} and \eqref{eq:contracting_pseudoscalar_tensor_with_q}.

\subsection{Hadronic tensor}
We start by using translational invariance of the vacuum to rewrite the hadronic tensor, \autoref{eq:definition_hadronic_tensor} as
\beq
    Q_B T_\had^{\mu \nu}(k,q) 
    = \int\dfx \ex^{\iu k x} \braket{0 | \tior\{J_\had^\nu(x) J_\elmag^\mu(0)\} | B^-}.
\eeq
By means of an integration by parts, a differentiation of the \LN{Heaviside} step function in the time-ordered product, and the \LN{Dirac} equation, we find
\beq\label{eq:contracting_hadronic_tensor_with_k}
    k_\nu [Q_B T_\had^{\mu \nu}(k,q)] 
    = Q_B T_\ps^\mu(k,q) + \iu \int \dff^3 x \, \ex^{-\iu \vec{k} \cdot \vec{x}} \braket{0 | [J_\had^0(\bar{x}),J_\elmag^\mu(0)] | B^-},
\eeq
where $\bar{x} = (x^0 = 0,\vec{x})$.
In the above, we furthermore used that a scalar--vector current--current matrix element of type $B$ meson to vacuum vanishes due to the involved quantum numbers, $\braket{0 | \tior\{J_S(x) J_\elmag^\mu(0)\} | B^-} = 0$, $J_S(x) = \bar{u}(x) b(x)$.
From an explicit calculation of the commutator in \autoref{eq:contracting_hadronic_tensor_with_k}, we finally arrive at
\beq
    k_\nu T_\had^{\mu \nu}(k,q) 
    = T_\ps^\mu(k,q) + f_B (k+q)^\mu,
\eeq
which is equivalent to \autoref{eq:contracting_homogeneous_hadronic_tensor_with_k} after inserting the decomposition of the hadronic tensor into its homogeneous and inhomogeneous parts, $T_\had^{\mu \nu}(k,q) = T_{\had,\hom}^{\mu \nu}(k,q) + T_{\had,\inhom}^{\mu \nu}(k,q)$, see \autoref{eq:hadronic_tensor_split}.

\subsection{Pseudoscalar tensor}
For the pseudoscalar tensor, we proceed similarly and use the definition in \autoref{eq:definition_pseudoscalar_tensor} to calculate
\beq
    q_\mu [Q_B T_\ps^\mu(k,q)] 
    = \iu \int \dff^3 x \, \ex^{-\iu \vec{q} \cdot \vec{x}} \braket{0 | [J_\elmag^0(\bar{x}),J_\ps(0)] | B^-}.
\eeq
An explicit calculation of the commutator results in \autoref{eq:contracting_pseudoscalar_tensor_with_q},
\beq
   q_\mu T_\ps^\mu(k,q) 
   = - f_B \mB^2.
\eeq

\section{\LN{Bardeen}--\LN{Tung}--\LN{Tarrach} procedure}
\label{app:BTT}
In this appendix, we outline the modification to the BTT procedure~\cite{Bardeen:1968ebo,Tarrach:1975tu} that leads us to the decomposition of the homogeneous part of the hadronic tensor into \LN{Lorentz} structures and form factors given in \autoref{eq:BTT_decomposition}.
To this end, we recall that the homogeneous part fulfills
\beq
    q_\mu T_{\had,\hom}^{\mu \nu}(k,q) 
    = 0,
\eeq
and that we additionally impose
\beq
    k_\nu T_{\had,\hom}^{\mu \nu}(k,q) 
    \overset{!}{=} T_{\ps,\hom}^\mu(k,q),
\eeq
see Eqs.~\eqref{eq:hadronic_tensor_split} and \eqref{eq:condition_pseudoscalar_homogeneous_part}, with $q_\mu T_{\ps,\hom}^\mu(k,q) = 0$.
Hence, we can split $T_{\had,\hom}^{\mu \nu}(k,q)$ according to
\beq\label{eq:homogeneous_tensor_split}
    T_{\had,\hom}^{\mu \nu}(k,q) 
    = \widetilde{T}_{\had,\hom}^{\mu \nu}(k,q) + T_{\ps,\hom}^\mu(k,q) \frac{k^\nu}{k^2},
\eeq
where $q_\mu \widetilde{T}_{\had,\hom}^{\mu \nu}(k,q) = k_\nu \widetilde{T}_{\had,\hom}^{\mu \nu}(k,q) = 0$.
In the above, $T_{\ps,\hom}^\mu(k,q)$ necessarily comes with a factor $k^\nu/k^2$ due to its pseudoscalar nature; \Lat{cf.} the fact that the spin-$0$ component of a spin-$1$ field is of timelike polarization.
Since the explicit $k^2$-pole attached to $T_{\ps,\hom}^\mu(k,q)$ is thus an inherent feature of the pseudoscalar contribution, it needs to be regularized either by a zero in the accompanying form factor or by a corresponding contribution within $\widetilde{T}_{\had,\hom}^{\mu \nu}(k,q)$.
We follow the latter approach: we perform the BTT procedure for $T_{\ps,\hom}^\mu(k,q)$ and $\widetilde{T}_{\had,\hom}^{\mu \nu}(k,q)$ separately, where we use the native blueprint for the former and a variant that introduces an explicit $k^2$-pole to cancel the aforementioned pole of the pseudoscalar contribution for the latter.

We first perform the BTT procedure for $T_{\ps,\hom}^{\mu \nu}(k,q)$, where the only available building blocks for the \LN{Lorentz} structures are
\beq
    \{ L_{\ps,\hom,i}^\mu \} 
    = \{ k^\mu, q^\mu \}
\eeq
and gauge invariance in the form $q_\mu T_{\ps,\hom}^\mu(k,q) = 0$ is imposed by means of
\begin{align}
    \big\{ \widetilde{L}_{\ps,\hom,i}^\mu \big\} 
    &= {\mathcal{I}^\mu}_\alpha \big\{ L_{\ps,\hom,i}^\alpha \big\},
    &
    \mathcal{I}^{\mu \nu} 
    &= g^{\mu \nu} - \frac{k^\mu q^\nu}{k \cdot q}.
\end{align}
The resulting set
\beq
    \big\{ \widetilde{L}_{\ps,\hom,i}^\mu \big\} 
    = \bigg\{ 0, q^\mu - \frac{q^2}{k \cdot q} k^\mu \bigg\}
\eeq
consists of a single nonvanishing structure with a pole in $(k \cdot q)$.
Following the regular procedure, this irreducible pole is to be eliminated by multiplying with $(k \cdot q)$, leading to the structure
\beq\label{eq:BTT_pseudoscalar_result}
    \widehat{L}_{\ps,\hom}^\mu 
    = (k \cdot q) q^\mu - q^2 k^\mu.
\eeq

To perform the BTT procedure for $\widetilde{T}_{\had,\hom}^{\mu \nu}(k,q)$, we note that the interaction is of the form $V-A$.
Hence, the available building blocks for the \LN{Lorentz} structures are given by
\beq
    \{ L_{\had,\hom,i}^{\mu \nu} \} 
    = \big\{ g^{\mu \nu}, k^\mu k^\nu, k^\mu q^\nu, q^\mu k^\nu, q^\mu q^\nu, \eps^{\mu \nu \alpha \beta} k_\rho q_\sigma \big\}
\eeq
and we impose $q_\mu \widetilde{T}_{\had,\hom}^{\mu \nu}(k,q) = k_\nu \widetilde{T}_{\had,\hom}^{\mu \nu}(k,q) = 0$ by means of
\begin{align}
    \big\{ \widetilde{L}_{\had,\hom,i}^{\mu \nu}\big\} 
    &= {\mathcal{I}^\mu}_\alpha \big\{L_{\had,\hom,i}^{\alpha \beta}\big\} {\widetilde{\mathcal{I}}_\beta}^{\phantom{\beta}\nu},
    &
    \widetilde{\mathcal{I}}^{\mu \nu} 
    &= g^{\mu \nu} - \frac{k^\mu k^\nu}{k^2}.
\end{align}
The resulting set
\beq
    \big\{\widetilde{L}_{\had,\hom,i}^{\mu \nu}\big\} 
    = \bigg\{ g^{\mu \nu} - \frac{k^\mu q^\nu}{k \cdot q}, 0, 0, 0, \frac{q^2}{k^2} k^\mu k^\nu - \frac{q^2}{k \cdot q} k^\mu q^\nu - \frac{k \cdot q}{k^2} q^\mu k^\nu + q^\mu q^\nu, \eps^{\mu \nu \rho \sigma} k_\rho q_\sigma \bigg\}
\eeq
contains structures with poles in $(k \cdot q)$ as well as $k^2$.
While we explicitly keep the $k^2$ poles, as mentioned above, we get rid of one of the two poles in $(k \cdot q)$ by following the regular procedure, \Lat{i.e.}, by taking an appropriate linear combination with nonsingular coefficients and multiplying the remaining pole by $(k \cdot q)$.
This leads to the minimal~\cite{Bardeen:1968ebo,Tarrach:1975tu} set
\begin{align}\label{eq:BTT_result}
    \big\{ \widehat{L}_{\had,\hom,i}^{\mu \nu} \big\} 
    &= \big\{(k \cdot q)\widetilde{L}_{\had,\hom,1}^{\mu \nu}, \widetilde{L}_{\had,\hom,5}^{\mu \nu} - q^2 \widetilde{L}_{\had,\hom,1}^{\mu \nu}, \widetilde{L}_{\had,\hom,6}^{\mu \nu}\big\} \notag \\
    &= \big\{ (k \cdot q) g^{\mu \nu} - k^\mu q^\nu, \frac{q^2}{k^2} k^\mu k^\nu - \frac{k \cdot q}{k^2} q^\mu k^\nu + q^\mu q^\nu - q^2 g^{\mu \nu}, \eps^{\mu \nu \rho \sigma} k_\rho q_\sigma \big\}.
\end{align}

Combining Eqs.~\eqref{eq:BTT_pseudoscalar_result} and \eqref{eq:BTT_result} with \autoref{eq:homogeneous_tensor_split}, the homogeneous part of the hadronic tensor thus takes the form given in \autoref{eq:BTT_decomposition}.\footnote{%
    Note that for the decay of an electrically neutral $B$ meson, as opposed to the case of a charged $B$ meson considered in this article, no inhomogeneous contribution, \autoref{eq:hadronic_tensor_split}, is present.
    As a consequence, in this scenario, the associated form factors are readily free of kinematic singularities in $k^2$ and $q^2$ as well as kinematic zeroes in $q^2$ but contain an explicit kinematic zero in $k^2$ due to the singularities in the structures.}

\section{Form factor projectors}
\label{app:projectors}
In this appendix, we collect the formulae for the projectors $\proj_i^{\mu \nu}(k,q)$ that fulfill ${\proj_i}_{\mu \nu}(k,q) T_\had^{\mu \nu}(k,q) = \F_i(k^2,q^2)$, $i=1,\ldots,4$, and ${\proj_i}_{\mu \nu}(k,q) T_\had^{\mu \nu}(k,q) = f_B/\mB$, $i=5,6$, for an arbitrary choice of basis for $T_\had^{\mu \nu}(k,q)$, as introduced in \autoref{sec:hadronic_tensor}~\cite{Mertig:1990an,Shtabovenko:2016sxi,Shtabovenko:2020gxv,Patel:2015tea}:
\begin{align}\label{eq:projectors}
	\frac{1}{\mB}\proj_1^{\mu \nu}(k,q) 
	&= \frac{k \cdot q}{2 [(k \cdot q)^2 - k^2 q^2]} g^{\mu \nu} 
	+ \frac{3 q^2 (k \cdot q)}{2 [(k \cdot q)^2 - k^2 q^2]^2} k^\mu k^\nu
	- \frac{(k \cdot q)^2 + 2 k^2 q^2}{2 [(k \cdot q)^2 - k^2 q^2]^2} k^\mu q^\nu
	- \frac{3 (k \cdot q)^2}{2 [(k \cdot q)^2 - k^2 q^2]^2} q^\mu k^\nu \notag \\
	&\quad + \frac{3 k^2 (k \cdot q)}{2 [(k \cdot q)^2 - k^2 q^2]^2} q^\mu q^\nu, \notag \\
	\frac{1}{\mB}\proj_2^{\mu \nu}(k,q) 
	&= \frac{k^2}{2 [(k \cdot q)^2 - k^2 q^2]} g^{\mu \nu} 
	+ \frac{2 (k \cdot q)^2 + k^2 q^2}{2 [(k \cdot q)^2 - k^2 q^2]^2} k^\mu k^\nu
	- \frac{3 k^2 (k \cdot q)}{2 [(k \cdot q)^2 - k^2 q^2]^2} k^\mu q^\nu
	- \frac{3 k^2 (k \cdot q)}{2 [(k \cdot q)^2 - k^2 q^2]^2} q^\mu k^\nu \notag \\
	&\quad + \frac{3 k^4}{2 [(k \cdot q)^2 - k^2 q^2]^2} q^\mu q^\nu, \notag \\
	\frac{1}{\mB}\proj_3^{\mu \nu}(k,q) 
	&= \frac{1}{(k \cdot q)^2 - k^2 q^2} k^\mu k^\nu
    - \frac{2 k^2}{[(k \cdot q)^2 - k^2 q^2] [2 (k \cdot q) + q^2]} q^\mu k^\nu
	- \frac{k^2}{[(k \cdot q)^2 - k ^2 q^2] [2 (k \cdot q) + q^2]} q^\mu q^\nu, \notag \\
	\frac{1}{\mB}\proj_4^{\mu \nu}(k,q) 
	&= -\frac{\iu}{2 [(k \cdot q)^2 - k^2 q^2]} \eps^{\mu \nu \rho \sigma} k_\rho q_\sigma, \notag \\
	\mB \proj_5^{\mu \nu}(k,q) 
	&= -\frac{k \cdot q}{(k \cdot q)^2 - k^2 q^2} q^\mu k^\nu + \frac{k^2}{(k \cdot q)^2 - k^2 q^2} q^\mu q^\nu, \notag \\
	\mB \proj_6^{\mu \nu}(k,q) 
	&= \frac{q^2}{(k \cdot q)^2 - k^2 q^2} q^\mu k^\nu - \frac{k \cdot q}{(k \cdot q)^2 - k^2 q^2} q^\mu q^\nu.
\end{align}
At this, an ambiguity is hidden in how to collect the terms of the inhomogeneous part into basis structures in \autoref{eq:full_hadronic_tensor}, since different such choices will lead to another set of projectors than the ones given above.
However, any difference $\overline{\proj}_i^{\mu \nu}(k,q)$ between two sets of valid projectors is at most of the form
\beq
    \overline{\proj}_i^{\mu \nu}(k,q) 
    = A_i q^\mu \big[ k^\nu [ (k \cdot q) + q^2 ] - q^\nu [ (k \cdot q) + k^2 ] \big]
\eeq
for $i=3,5,6$, with some coefficient $A_i \equiv A_i(k^2,q^2)$, so that
\begin{align}
    {\overline{\proj}_i}_{\mu \nu}(k,q) T_{\had,\hom}^{\mu \nu}(k,q) 
    &= A_i \big[ q_\mu T_{\had,\hom}^{\mu \nu}(k,q) \big] \big[ k_\nu [ (k \cdot q) + q^2 ] - q_\nu [ (k \cdot q) + k^2 ] \big] \notag \\
    &= 0, \notag \\
    {\overline{\proj}_i}_{\mu \nu}(k,q) T_{\had,\inhom}^{\mu \nu}(k,q)
    &= A_i \big[ q_\mu T_{\had,\inhom}^{\mu \nu}(k,q) \big] \big[ k_\nu [ (k \cdot q) + q^2 ] - q_\nu [ (k \cdot q) + k^2 ] \big] \notag \\
    &= A_i [ -f_B (k+q)^\nu ] \big[ k_\nu [ (k \cdot q) + q^2 ] - q_\nu [ (k \cdot q) + k^2 ] \big] \notag \\
    &= 0.
\end{align}
For $i=1,2,4$, the projectors are independent of this choice, \Lat{i.e.}, $A_i = 0$.

\section{Kinematics}
\label{app:kinematics}
In this appendix, we present some details on the kinematics for the processes $B^- \to \ell^- \bar{\nu}_\ell \gamma^*$ and $B^- \to \ell^- \bar{\nu}_\ell \ellp \ellpbar$, which are necessary ingredients to calculate the squared spin-averaged amplitudes $\lvert \overline{\M}(B^- \to \ell^- \bar{\nu}_\ell \gamma^*) \rvert^2$ in \autoref{eq:M_B_to_l_nu_gamma_squared} and $\lvert \overline{\M}(B^- \to \ell^- \bar{\nu}_\ell \ellp \ellpbar) \rvert^2$ in \autoref{sec:pheno}.

\subsection{\boldmath{$B^- \to \ell^- \bar{\nu}_\ell \gamma^*$}}
For a consistent treatment of the kinematics in the process $B^- \to \ell^- \bar{\nu}_\ell \gamma^*$, all momenta and polarization vectors have to be evaluated in a single frame of reference.
To this end, we calculate the corresponding quantities in the CMS of the $\{\ell^- \bar{\nu}_\ell(k),\gamma^*(q)\}$ and $\{\ell^-(p_\ell),\bar{\nu}_\ell(p_\nu)\}$ subsystem and perform a \LN{Lorentz} transformation of the latter to the former frame.

In the CMS $\{\ell^- \bar{\nu}_\ell(k),\gamma^*(q)\}$, one finds the magnitude of the photon's three-momentum and the energies
\begin{align}
    \lvert \vec{p}_\gamma \rvert 
    &= \frac{\sqrt{\lambda(\mB^2,k^2,q^2)}}{2\mB},
    &
    E_{\ell \nu} 
    &= \frac{\mB^2 + k^2 -q^2}{2\mB},
    &
    E_\gamma 
    &= \frac{\mB^2 - k^2 + q^2}{2\mB}.
\end{align}
The four-momentum of the leptonic subsystem thus reads
\beq
    k 
    = (E_{\ell \nu},0,0,\lvert \vec{p}_\gamma \rvert)^\intercal
\eeq
and, accordingly, the four-momentum of the photon and its polarization vectors are given by
\begin{align}
    q 
    &= (E_\gamma,0,0,-\lvert \vec{p}_\gamma \rvert)^\intercal,
    &
    \eps(q; \lambda = \pm 1) 
    &= \mp \frac{1}{\sqrt{2}}(0,1,\mp \iu,0)^\intercal, \notag
    \\
    \eps(q; \lambda = 0) 
    &= \frac{1}{\xi}(-\lvert \vec{p}_\gamma \rvert,0,0,E_\gamma)^\intercal,
    &
    \eps(q; \lambda = T) 
    &= \frac{1}{\xi}(E_\gamma,0,0,-\lvert \vec{p}_\gamma \rvert)^\intercal,
\end{align}
where any physical observable necessarily needs to be independent of $\xi = \sqrt{q^2}$.

In the CMS $\{\ell^-(p_\ell),\bar{\nu}_\ell(p_\nu)\}$, we have
\begin{align}
    \lvert \vec{p}_\ell \rvert 
    &= \frac{k^2 - m_\ell^2}{2\sqrt{k^2}},
    &
    E_\ell 
    &= \frac{k^2 + m_\ell^2}{2\sqrt{k^2}},
    &
    E_\nu 
    &= \frac{k^2 - m_\ell^2}{2\sqrt{k^2}}
\end{align}
for the magnitude of the negatively charged lepton's three-momentum and the corresponding energies.
Hence, transforming the subsystem $\{\ell^-(p_\ell),\bar{\nu}_\ell(p_\nu)\}$ to the CMS $\{\ell^- \bar{\nu}_\ell(k),\gamma^*(q)\}$, the four-momenta of the leptons are found to be
\begin{align}
    p_\ell 
    &= 
    \begin{pmatrix}
        \upgamma_{\ell \nu, \gamma}(E_\ell + \beta_{\ell \nu, \gamma} \lvert \vec{p}_\ell \rvert \cos\vartheta_W) \\
        \lvert \vec{p}_\ell \rvert \sin\vartheta_W \\
        0 \\
        \upgamma_{\ell \nu, \gamma}(\beta_{\ell \nu, \gamma} E_\ell + \lvert \vec{p}_\ell \rvert \cos\vartheta_W)
    \end{pmatrix},
    &
    p_\nu 
    &= 
    \begin{pmatrix}
        \upgamma_{\ell \nu, \gamma}(E_\nu - \beta_{\ell \nu, \gamma} \lvert \vec{p}_\ell \rvert \cos\vartheta_W) \\
        -\lvert \vec{p}_\ell \rvert \sin\vartheta_W \\
        0 \\
        \upgamma_{\ell \nu, \gamma}(\beta_{\ell \nu, \gamma} E_\nu - \lvert \vec{p}_\ell \rvert \cos\vartheta_W)
    \end{pmatrix},
\end{align}
where $\beta_{\ell \nu, \gamma} = \lvert \vec{p}_\gamma \rvert/E_{\ell \nu}$, $\upgamma_{\ell \nu, \gamma} = (1-\beta_{\ell \nu, \gamma}^2)^{-1/2}$, and $\vartheta_W$ is the polar angle of $\ell^-(p_\ell)$ in the CMS $\{\ell^-(p_\ell),\bar{\nu}_\ell(p_\nu)\}$.

\subsection{\boldmath{$B^- \to \ell^- \bar{\nu}_\ell \ellp \ellpbar$}}
\begin{figure}[t]
    \centering
    \includegraphics[scale=0.625]{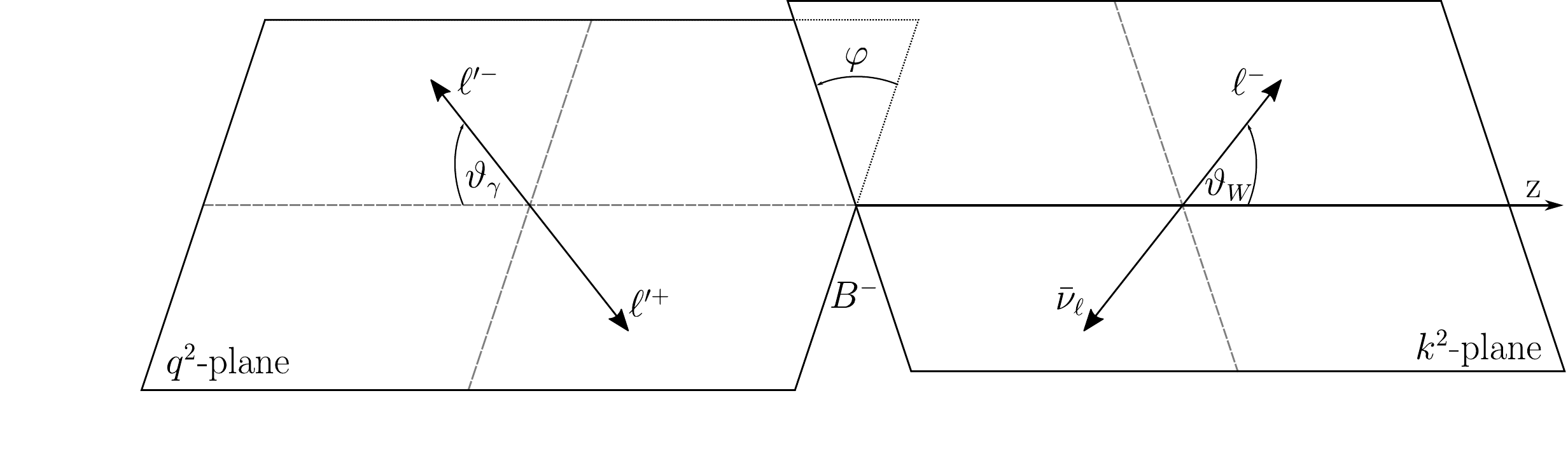}
    \caption{Illustration of the decay $B^- \to \ell^- \bar{\nu}_\ell \ellp \ellpbar$ along with the two decay planes of the leptonic subsystems and the three angles necessary to describe the kinematics of the process.}
    \label{fig:illustration_kinematics_B_to_3l_nu}
\end{figure}

In addition to the magnitudes of three-momenta $\lvert \vec{p}_\gamma \rvert$ and $\lvert \vec{p}_\ell \rvert$ in the CMS $\{\ell^- \bar{\nu}_\ell(k),\gamma^*(q)\}$ and $\{\ell^-(p_\ell),\bar{\nu}_\ell(p_\nu)\}$, respectively, we need the three-momentum of $\ellp(q_1)$ in the CMS $\{\ellp(q_1), \ellpbar(q_2)\}$ to describe the kinematics of the process $B^- \to \ell^- \bar{\nu}_\ell \ellp \ellpbar$,
\beq
    \lvert \vec{p}_{\ell'} \rvert 
    = \frac{\sqrt{q^2 - 4m_{\ell'}^2}}{2}.
\eeq
Furthermore, two additional angles besides $\vartheta_W$ are necessary here: the polar angle $\vartheta_\gamma$ of $\ellp(q_1)$ in the CMS $\{\ellp(q_1), \ellpbar(q_2)\}$ and the azimuthal angle $\varphi$ between the decay planes of the subsystems $\{\ell^-(p_\ell),\bar{\nu}_\ell(p_\nu)\}$ and $\{\ellp(q_1), \ellpbar(q_2)\}$, see \autoref{fig:illustration_kinematics_B_to_3l_nu}.

For the four-body phase space, we used
\beq
    \dff \Phi_4(p; p_\ell,p_\nu,q_1,q_2) 
    = \dff \Phi_2(p;k,q) \dff \Phi_2(k;p_\ell,p_\nu) \dff \Phi_2(q;q_1,q_2) \frac{\dff k^2}{2\pi} \frac{\dff q^2}{2\pi}
\eeq
in \autoref{eq:four_body_phase_space}, where
\begin{align}
    \dff \Phi_2(p;k,q) 
    &= \frac{1}{16 \pi^2} \frac{\lvert \vec{p}_\gamma \rvert}{\mB} \dff \Omega_B,
    &
    \dff \Phi_2(k;p_\ell,p_\nu) 
    &= \frac{1}{16 \pi^2} \frac{\lvert \vec{p}_\ell \rvert}{\sqrt{k^2}} \dff \Omega_W,
    &
    \dff \Phi_2(q;q_1,q_2) 
    &= \frac{1}{16 \pi^2} \frac{\lvert \vec{p}_{\ell'} \rvert}{\sqrt{q^2}} \dff \Omega_\gamma
\end{align}
are the two-body phase spaces of the subsystems $\{\ell^- \bar{\nu}_\ell(k),\gamma^*(q)\}$, $\{\ell^-(p_\ell),\bar{\nu}_\ell(p_\nu)\}$, and $\{\ellp(q_1), \ellpbar(q_2)\}$, respectively.
Here, $\dff \Omega_B$, $\dff \Omega_W$, and $\dff \Omega_\gamma$ denote the differential solid angles in the corresponding CMS.
Three of the six angular integrations can be rendered trivial to carry out by rotating the coordinate system appropriately, leading to the expression
\beq
    \dff \Phi_4(p; p_\ell,p_\nu,q_1,q_2) 
    = \frac{1}{2048 \pi^6} \frac{\lvert \vec{p}_\gamma \rvert}{\mB} \frac{\lvert \vec{p}_{\ell} \rvert}{\sqrt{k^2}} \frac{\lvert \vec{p}_{\ell'} \rvert}{\sqrt{q^2}} \dff \cos\vartheta_{W} \dff \cos\vartheta_{\gamma} \dff \varphi \, \dff k^2 \dff q^2
\eeq
for the four-body phase space, with the remaining angles being as illustrated in \autoref{fig:illustration_kinematics_B_to_3l_nu}.

\section{Asymptotics}
\label{app:asymptotics}
\begin{figure}[t]
    \centering
    \includegraphics[scale=0.875]{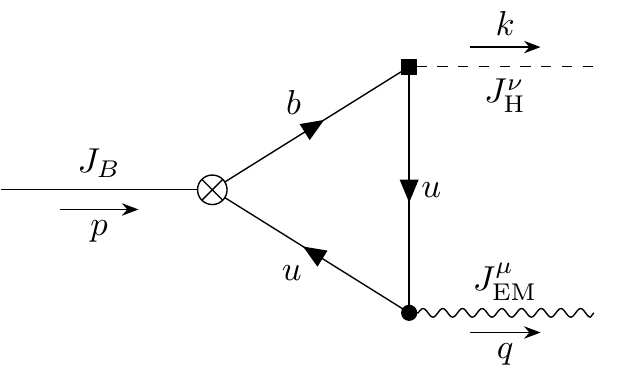}
    \qquad
    \qquad
    \qquad
    \qquad
    \includegraphics[scale=0.875]{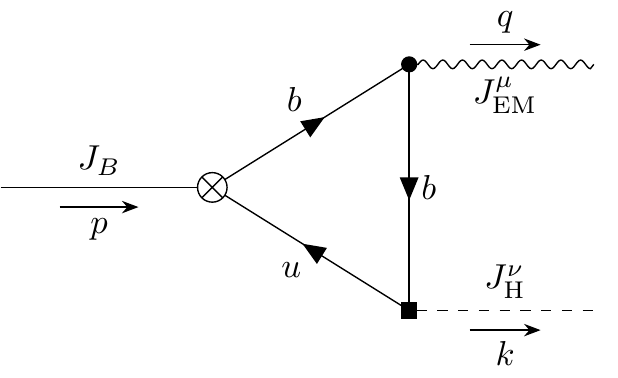}
    \caption{The leading-order diagrams in the OPE for the form factors $\F_i(k^2,q^2)$.
    Diagrams contributing at a higher order in the OPE are neglected here.}
    \label{fig:Subtractions_OPE}
\end{figure}

In this appendix, we show that the form factors $\F_i^I(k^2,q^2)$ introduced in \autoref{sec:dispersion_relations} as well as their discontinuities drop off as $1/q^2$ asymptotically.
This behavior was assumed to avoid subtracting the dispersion relation of \autoref{eq:cauchys_integral_formula} and justified the monopole-like ansatz for the form factors in \autoref{eq:form_factors_series_coefficients}.
We determine the form factors' asymptotic behavior for $q^2 \to \infty$ by inspecting the results of a calculation of the $B\to \gamma$ form factors within an operator product expansion (OPE)~\cite{Janowski:2021yvz}.
For our purposes, it suffices to inspect the leading-power terms within this OPE, which are diagrammatically depicted in \autoref{fig:Subtractions_OPE}.
The OPE uses an interpolating quark current for the $B$ meson, namely~\cite{Janowski:2021yvz} $J_B(x) = \bar{u}(x) \gamma_5 b(x)$, which fulfills $\braket{0 | J_B(0) | B^-} = -\iu \mB^2 f_B/(m_b + m_u)$.
We then calculate the sum of the two diagrams depicted in \autoref{fig:Subtractions_OPE}, leading to
\begin{align}
    X_{\mu \nu}^{I}(k,q) &= e \int \frac{\dff^4 l}{(2\pi)^4} \Tr\bigg[ -\gamma_5 \frac{\iu (\slashed{l} - \slashed{q} + m_u)}{(l - q)^2 - m_u^2} Q_u^I \gamma_\mu \frac{\iu (\slashed{l} + m_u)}{l^2 - m_u^2} \gamma_\nu (1 - \gamma_5) \frac{\iu (\slashed{l} + \slashed{k} + m_b)}{(l + k)^2 - m_b^2} \notag \\
    &\qquad \qquad \qquad \quad
    - \gamma_5 \frac{\iu (\slashed{l} - \slashed{k} + m_u)}{(l - k)^2 - m_u^2} \gamma_\nu (1 - \gamma_5) \frac{\iu (\slashed{l} + m_b)}{l^2 - m_b^2} Q_b^I \gamma_\mu \frac{\iu (\slashed{l} + \slashed{q} + m_b)}{(l + q)^2 - m_b^2} \bigg],
\end{align}
where $l$ is the loop momentum and $q^2 < 0$ large.
The isospin charges are given by $(Q_u^{I=0},Q_b^{I=0}) = (1/6,-1/3)$ and $(Q_u^{I=1},Q_b^{I=1}) = (1/2,0)$.

For the discontinuities, it then follows that
\beq
    \discq\F_i^{I,\OPE}(k^2,q^2) \propto \discq \big[ {\proj_i}^{\mu \nu}(k,q) X_{\mu \nu}^I(k,q) \big],
\eeq
so that the asymptotic behavior for large $q^2 < 0$ is found to be given by~\cite{Patel:2015tea}
\beq
    \discq\F_{i}^{I,\OPE}(k^2,q^2) \sim 1/q^2,
\eeq
rendering the dispersion integrals convergent without any subtractions.

Similarly, we find
\beq
    \F_i^{I,\OPE}(k^2,q^2) \sim 1/q^2
\eeq
for the asymptotic behavior of the form factors, so that a monopole-like ansatz in the framework of VMD is justified.

\section{Intermediate results}
\label{app:intermediate_results}
In this appendix, we collect the covariance matrices for the normalizations $N_{i,j}^V$ from \autoref{tab:z_expansion_form_factors} and the functions $f_{i,j}$ as well as $g_{i,j}$ introduced in Eqs.~\eqref{eq:two_fold_decay_width} and \eqref{eq:FB_asymmetry}.

\subsection{Covariance matrices}
For reasons of consistency with the rounding of the uncertainties on the normalizations, we round the numerical values in the covariance matrices to four significant digits.
Because of the fact that the input used to determine the normalizations does not exhibit a correlation between the parameters of the $\omega$ and $\rho$, the normalizations $N_{i,j}^\omega$ and $N_{i,j}^\rho$ are uncorrelated, \Lat{i.e.}, $\Cov(N_{i,j}^\omega,N_{k,l}^\rho) = 0$ for all $i,j,k,l$, so that our results can be collected in two $(12 \times 12)$ matrices.

For the covariances between the normalizations $N_{i,j}^\omega$, we find
\begin{align}
    &10^6 \times \Cov(N_{i,j}^\omega,N_{k,l}^\omega)_{m n} = \\
    &
    \begin{pmatrix}
        9.186 & -11.29 & 66.84 & 16.05 & -65.26 & 739.2 & 16.05 & 57.91 & -371.5 & -7.348 & -37.43 & 135.1 \\
        -11.29 & 378.7 & -1444 & -151.5 & 491.6 & -3209 & -151.5 & -186.7 & 1270 & 2.220 & -241.8 & 353.4 \\
        66.84 & -1444 & 7180 & 991.8 & -5858 & 24670 & 991.8 & -911.2 & -1778 & -6.404 & 558.3 & -1611 \\
        16.05 & -151.5 & 991.8 & 740.4 & -1134 & 7731 & 740.4 & 2429 & -6410 & 13.70 & 3.901 & 134.0 \\
        -65.26 & 491.6 & -5858 & -1134 & 20370 & -52440 & -1134 & 14440 & -31960 & -9.187 & 986.6 & -2592 \\
        739.2 & -3209 & 24670 & 7731 & -52440 & 266600 & 7731 & -9322 & 46070 & -305.2 & -3351 & 16080 \\
        16.05 & -151.5 & 991.8 & 740.4 & -1134 & 7731 & 740.4 & 2429 & -6410 & 13.70 & 3.901 & 134.0 \\
        57.91 & -186.7 & -911.2 & 2429 & 14440 & -9322 & 2429 & 28910 & -63340 & 15.82 & 682.5 & 204.0 \\
        -371.5 & 1270 & -1778 & -6410 & -31960 & 46070 & -6410 & -63340 & 498000 & 144.6 & -1346 & 13990 \\
        -7.348 & 2.220 & -6.404 & 13.70 & -9.187 & -305.2 & 13.70 & 15.82 & 144.6 & 7.794 & 34.85 & -134.1 \\
        -37.43 & -241.8 & 558.3 & 3.901 & 986.6 & -3351 & 3.901 & 682.5 & -1346 & 34.85 & 450.4 & -1108 \\
        135.1 & 353.4 & -1611 & 134.0 & -2592 & 16080 & 134.0 & 204.0 & 13990 & -134.1 & -1108 & 8249
    \end{pmatrix}, \notag
\end{align}
where $m=(3i + j - 2)$ and $n=(3k + l - 2)$ denote the rows and columns of the matrix, respectively.
At this, it is to be noted that $N_{2,0}^\omega = N_{3,0}^\omega$, see the discussion in \autoref{sec:dispersion_relations}, so that one row and one column of the matrix is, in fact, redundant, reducing the degrees of freedom to an $(11 \times 11)$ matrix.

In the same way and with the analogous caveat $N_{2,0}^\rho = N_{3,0}^\rho$, we find the covariances between the normalizations $N_{i,j}^\rho$ to be given by
\begin{align}
    &10^5 \times \Cov(N_{i,j}^\rho,N_{k,l}^\rho)_{m n} = \\
    &
    \begin{pmatrix}
        7.758 & -25.35 & 132.8 & 17.88 & -47.77 & 705.8 & 17.88 & 70.38 & -233.2 & -5.403 & -18.46 & 46.61 \\
        -25.35 & 231.1 & -988.0 & -151.1 & 393.9 & -3906 & -151.1 & -389.7 & 7.717 & 12.95 & -55.78 & -26.32 \\
        132.8 & -988.0 & 5543 & 1059 & -5304 & 26800 & 1059 & 411.8 & -3620 & -55.19 & -90.41 & -25.79 \\
        17.88 & -151.1 & 1059 & 631.5 & -1626 & 7978 & 631.5 & 1294 & -2278 & 9.762 & -23.75 & -12.89 \\
        -47.77 & 393.9 & -5304 & -1626 & 17200 & -43390 & -1626 & 7476 & -12080 & -44.36 & 814.8 & -1798 \\
        705.8 & -3906 & 26800 & 7978 & -43390 & 224000 & 7978 & 2795 & 15740 & -248.9 & -2166 & 8948 \\
        17.88 & -151.1 & 1059 & 631.5 & -1626 & 7978 & 631.5 & 1294 & -2278 & 9.762 & -23.75 & -12.89 \\
        70.38 & -389.7 & 411.8 & 1294 & 7476 & 2795 & 1294 & 14980 & -16040 & -29.65 & 443.1 & 795.0 \\
        -233.2 & 7.717 & -3620 & -2278 & -12080 & 15740 & -2278 & -16040 & 396500 & 42.62 & 123.5 & 27430 \\
        -5.403 & 12.95 & -55.19 & 9.762 & -44.36 & -248.9 & 9.762 & -29.65 & 42.62 & 5.693 & 16.05 & -56.63 \\
        -18.46 & -55.78 & -90.41 & -23.75 & 814.8 & -2166 & -23.75 & 443.1 & 123.5 & 16.05 & 197.2 & -383.9 \\
        46.61 & -26.32 & -25.79 & -12.89 & -1798 & 8948 & -12.89 & 795.0 & 27430 & -56.63 & -383.9 & 5497
    \end{pmatrix}. \notag
\end{align}

\subsection{Functions $f_{i,j}$ and $g_{i,j}$}
For the functions $f_{i,j}$ introduced in \autoref{eq:two_fold_decay_width}, we obtain
\begin{align}
    f_{1,1} &= \frac{64\pi k_-^2 q_+^2 (k_+^2 + k^2) [\lambda(\mB^2,k^2,q^2) + 6k^2 q^2]}{9k^2},
    &
    f_{2,2} &= \frac{32\pi k_-^2 q_+^2 q^2 (k_+^2 + k^2) [\lambda(\mB^2,k^2,q^2) + 12k^2 q^2]}{9k^4}, \notag \\
    f_{3,3} &= m_\ell^2 \frac{32\pi k_-^2 q_+^2 q^2 \lambda(\mB^2,k^2,q^2)}{3k^4}, 
    &
    f_{4,4} &= \frac{64\pi k_-^2 q_+^2 (k_+^2 + k^2) \lambda(\mB^2,k^2,q^2)}{9k^2}, \notag \\
    f_{1,2} &= -\frac{64\pi k_-^2 q_+^2 q^2 (k_+^2 + k^2) \Delta(k^2,q^2)}{3k^2}
\end{align}
and
\begin{align}
    f_{1,5} &= m_\ell^2 \frac{128\pi q_+^2}{3k_-^2} \Big[ k_-^2 \big[ \Delta(k^2,q^2) - k_+^2 \big] - k_B^2 k^2 \big[ \Delta(k^2,q^2) - 2m_\ell^2 \big] L_D(k^2,q^2) \Big], \\
    f_{2,5} &= -m_\ell^2 \frac{128\pi q_+^2 q^2}{3k_-^2} \Big[ 3k_-^2 - \big[ 3k_B^2 k^2 + (k_-^2)^2 \big] L_D(k^2,q^2) \Big], \notag \\
    f_{3,5} &= -m_\ell^2 \frac{64\pi q_+^2 q^2}{3k_B^2 k_-^2 k^2} \Big[ k_-^2 \big[ k_-^2 \Delta(k^2,q^2) + 2k^2 (k_B^2 + 2k_-^2) \big] - 2 k_B^2 k^2 (k_B^2 k^2 + k_-^2 k_+^2) L_D(k^2,q^2) \Big], \notag \\
    f_{4,5} &= -m_\ell^2 \frac{128\pi q_+^2}{3k_-^2} \Big[ k_-^2 \Delta(k^2,q^2) - k^2 \big[ k_B^2 \Delta(k^2,q^2) - 2k_-^2 q^2 \big] L_D(k^2,q^2) \Big], \notag \\
    f_{5,5} &= -m_\ell^2 \frac{128\pi q_+^2}{3(k_B^2)^2 k_-^2 [k_-^2 q^2 (k_B^2 + k_-^2) + m_\ell^2 (k_B^2)^2]} \notag \\
    &\quad \times \Bigg[ k_-^2 \bigg[ k_-^2 (k_B^2 + k_-^2) \Big[ k_B^2 - \Delta(k^2,q^2) \Big] \Big[ 4(k_-^2)^2 + k_-^2 \big[ 3k_B^2 + \Delta(k^2,q^2) \big] + 4(k_B^2)^2 \Big] \notag \\
    &\qquad\qquad + m_\ell^2 \Big[ 4(k_-^2)^3 \big[ k_B^2 - \Delta(k^2,q^2) \big] + 8k_B^2 (k_-^2)^2 \big[ 2k_B^2 - \Delta(k^2,q^2) \big] + (k_B^2)^2 k_-^2 \big[ 13k_B^2 - 5\Delta(k^2,q^2) \big] \notag \\
    &\qquad\qquad\qquad\quad + 2(k_B^2)^3 \big[ 2k_B^2 - \Delta(k^2,q^2) \big] \Big] + 8m_\ell^4 (k_B^2)^2 (k_B^2 + k_-^2) \bigg] \notag \\
    &\qquad\quad + 2k_B^2 k^2 \Big[ k_B^2 \big[ \Delta(k^2,q^2) - 2k_B^2 \big] - 2k_-^2 (k_B^2 + k_-^2) - 4m_\ell^2 (k_B^2 + k_-^2) \Big] \big[ k_-^2 q^2 (k_B^2 + k_-^2) + m_\ell^2 (k_B^2)^2 \big] L_D(k^2,q^2) \Bigg], \notag
\end{align}
where we defined
\begin{align}
    k_B^2 &= \mB^2 - k^2,
    &
    k_\pm^2 &= k^2 \pm m_\ell^2,
    &
    q_+^2 &= q^2 + 2m_{\ell'}^2, \notag \\
    \Delta(k^2,q^2) &= k_B^2 - q^2,
    &
    L_D &= \frac{L_+(k^2,q^2) - L_-(k^2,q^2)}{\sqrt{\lambda(\mB^2,k^2,q^2)}},
    &
    L_\pm(k^2,q^2) &= \log\bigg[1 \pm \frac{k_-^2 \sqrt{\lambda(\mB^2,k^2,q^2)}}{k_B^2 k_+^2 + k_-^2 q^2}\bigg].
\end{align}
All other, unlisted functions vanish, \Lat{i.e.}, $f_{1,3} = f_{1,4} = f_{2,3} = f_{2,4} = f_{3,4} = 0$.
Given the scaling with the lepton mass, one finds that this set further reduces to four functions in the chiral limit $m_\ell = 0$.

For the functions $g_{i,j}$ introduced in \autoref{eq:FB_asymmetry}, we similarly obtain
\begin{align}
    g_{1,3} &= m_\ell^2 \frac{32\pi k_-^2 q_+^2 q^2 \sqrt{\lambda(\mB^2,k^2,q^2)}}{3k^2}, 
    &
    g_{1,4} &= \frac{32\pi k_-^2 q_+^2 \Delta(k^2,q^2) \sqrt{\lambda(\mB^2,k^2,q^2)}}{3}, \notag \\
    g_{2,3} &= -m_\ell^2 \frac{16\pi k_-^2 q_+^2 q^2 \Delta(k^2,q^2) \sqrt{\lambda(\mB^2,k^2,q^2)}}{3k^4},
    &
    g_{2,4} &= -\frac{64\pi k_-^2 q_+^2 q^2 \sqrt{\lambda(\mB^2,k^2,q^2)}}{3}
\end{align}
and
\begin{align}
    g_{1,5} &= m_\ell^2 \frac{32\pi q_+^2}{3k_B^2 k_-^2} \Big[ 2(k_-^2)^2 \sqrt{\lambda(\mB^2,k^2,q^2)} - 4(k_B^2)^2 k^2 \big[ \Delta(k^2,q^2) - 2m_\ell^2 \big] \widetilde{L}_D(k^2,q^2) \Big], \\
    g_{2,5} &= m_\ell^2 \frac{32\pi q_+^2 q^2}{3k_B^2 k_-^2 k^2} \Big[ (k_-^2)^2 \sqrt{\lambda(\mB^2,k^2,q^2)} + 4k_B^2 k^2 \big[ 3k_B^2 k^2 + (k_-^2)^2 \big] \widetilde{L}_D(k^2,q^2) \Big], \notag \\
    g_{3,5} &= m_\ell^2 \frac{128\pi q_+^2 q^2}{3k_-^2} (k_B^2 k^2 + k_+^2 k_-^2) \widetilde{L}_D(k^2,q^2), \notag \\
    g_{4,5} &= m_\ell^2 \frac{128\pi k^2 q_+^2}{3k_-^2} \Big[ 2k_-^2 \big[ \Delta(k^2,q^2) - k_B^2 \big] + k_B^2 \Delta(k^2,q^2) \Big] \widetilde{L}_D(k^2,q^2), \notag \\
    g_{5,5} &= m_\ell^2 \frac{256\pi k^2 q_+^2}{3k_B^2 k_-^2 [k_-^2 q^2 (k_B^2 + k_-^2) + m_\ell^2 (k_B^2)^2][k_+^2 \Delta(k^2,q^2) + 2k^2 q^2]} \notag \\
    &\quad \times \bigg[ k_B^2 (k_-^2)^2 (k_B^2 + k_-^2) (q^2 + 2m_\ell^2) \sqrt{\lambda(\mB^2,k^2,q^2)} + \big[ k_+^2 \Delta(k^2,q^2) + 2k^2 q^2 \big] \big[ m_\ell^2 (k_B^2)^2 + k_-^2 q^2 (k_B^2 + k_-^2) \big] \notag \\
    &\qquad\qquad\qquad\qquad\qquad\qquad\qquad\qquad\quad \times \Big[ 4m_\ell^2 (k_B^2 + k_-^2) + 2 \big[ (k_B^2)^2 + (k_-^2)^2 + k_B^2 k_-^2 \big] - k_B^2 \Delta(k^2,q^2) \Big] \widetilde{L}_D(k^2,q^2) \bigg], \notag
\end{align}
where we additionally defined
\begin{equation}
    \widetilde{L}_D = \frac{1}{\sqrt{\lambda(\mB^2,k^2,q^2)}} \log\left[ \frac{4k_-^2 k^2 q^2 (k_B^2 + k_-^2) + 4m_\ell^2 (k_B^2)^2 k^2}{[k_+^2 \Delta(k^2,q^2) + 2k^2 q^2]^2} \right].
\end{equation}
All other, unlisted functions vanish, \Lat{i.e.}, $g_{1,1} = g_{2,2} = g_{3,3} = g_{4,4} = g_{1,2} = g_{3,4} = 0$.
Again, from the scaling with the lepton mass, one finds that this set further reduces to two functions in the chiral limit $m_\ell = 0$.

\section{Constants and parameters}
\label{app:constants}
We collect the constants and parameters used throughout our analysis in \autoref{tab:constants_and_parameters}.
\begin{table}[t]
	\centering
	\begin{tabular}{c  c  c  c}
	\toprule
		Quantity & Variable & Value & Reference \\
		\midrule
		Mass $\pi^\pm$ & $\Mpi$ & $139.57039(18) \MeV$ & \cite{ParticleDataGroup:2020ssz} \\
		Mass $B^\pm$ & $\mB$ & $5279.34(12) \MeV$ & \cite{ParticleDataGroup:2020ssz} \\
		Mass $B^*$ & $\mBStar$ & $5324.71(21) \MeV$ & \cite{ParticleDataGroup:2020ssz} \\
		Mass $B_1$ & $\mBOne$ & $5725.9^{+2.5}_{-2.7} \MeV$ & \cite{ParticleDataGroup:2020ssz} \\
		Mass $\rho^0(770)$ & $\Mrho$ & $775.26(23) \MeV$ & \cite{ParticleDataGroup:2020ssz} \\
		Mass $\omega(782)$ & $\Momega$ & $782.66(13) \MeV$ & \cite{ParticleDataGroup:2020ssz} \\
		Lifetime $B^\pm$ & $\tau_B$ & $1638(4) \, \text{fs}$ & \cite{ParticleDataGroup:2020ssz} \\
		Width $\rho^0(770)$ & $\Gamma_\rho$ & $147.4(8) \MeV$ & \cite{ParticleDataGroup:2020ssz} \\	
		Width $\omega(782)$ & $\Gamma_\omega$ & $8.68(13) \MeV$ & \cite{ParticleDataGroup:2020ssz} \\
		%
		\midrule
		Decay constant $\rho^0(770)$ & $f_\rho$ & $216(3) \MeV$ & \cite{Bharucha:2015bzk} \\
		Decay constant $\omega(782)$ & $f_\omega$ & $197(8) \MeV$ & \cite{Bharucha:2015bzk} \\
		%
		\midrule
        Decay constant $B^\pm$ & $f_B$ & $190.0(1.3) \MeV$ & \cite{Aoki:2021kgd,TUMQCD:2018fsq,ETM:2016nbo,Dowdall:2013tga,Hughes:2017spc} \\	
		%
		\midrule
        CKM matrix element $b \to u$ & $\lvert V_{ub} \rvert$ & $3.77(15) \times 10^{-3}$ & \cite{Leljak:2021vte} \\
        \bottomrule
	\end{tabular}
	\caption{The masses, widths, and other physical parameters needed for the calculations in this article.}
	\label{tab:constants_and_parameters}
\end{table}

\bibliographystyle{apsrev4-2}

\bibliography{Bib/bibliography} 

\begin{thebibliography}{43}%
\makeatletter
\providecommand \@ifxundefined [1]{%
 \@ifx{#1\undefined}
}%
\providecommand \@ifnum [1]{%
 \ifnum #1\expandafter \@firstoftwo
 \else \expandafter \@secondoftwo
 \fi
}%
\providecommand \@ifx [1]{%
 \ifx #1\expandafter \@firstoftwo
 \else \expandafter \@secondoftwo
 \fi
}%
\providecommand \natexlab [1]{#1}%
\providecommand \enquote  [1]{``#1''}%
\providecommand \bibnamefont  [1]{#1}%
\providecommand \bibfnamefont [1]{#1}%
\providecommand \citenamefont [1]{#1}%
\providecommand \href@noop [0]{\@secondoftwo}%
\providecommand \href [0]{\begingroup \@sanitize@url \@href}%
\providecommand \@href[1]{\@@startlink{#1}\@@href}%
\providecommand \@@href[1]{\endgroup#1\@@endlink}%
\providecommand \@sanitize@url [0]{\catcode `\\12\catcode `\$12\catcode
  `\&12\catcode `\#12\catcode `\^12\catcode `\_12\catcode `\%12\relax}%
\providecommand \@@startlink[1]{}%
\providecommand \@@endlink[0]{}%
\providecommand \url  [0]{\begingroup\@sanitize@url \@url }%
\providecommand \@url [1]{\endgroup\@href {#1}{\urlprefix }}%
\providecommand \urlprefix  [0]{URL }%
\providecommand \Eprint [0]{\href }%
\providecommand \doibase [0]{https://doi.org/}%
\providecommand \selectlanguage [0]{\@gobble}%
\providecommand \bibinfo  [0]{\@secondoftwo}%
\providecommand \bibfield  [0]{\@secondoftwo}%
\providecommand \translation [1]{[#1]}%
\providecommand \BibitemOpen [0]{}%
\providecommand \bibitemStop [0]{}%
\providecommand \bibitemNoStop [0]{.\EOS\space}%
\providecommand \EOS [0]{\spacefactor3000\relax}%
\providecommand \BibitemShut  [1]{\csname bibitem#1\endcsname}%
\let\auto@bib@innerbib\@empty
\bibitem [{\citenamefont {Beneke}\ and\ \citenamefont
  {Rohrwild}(2011)}]{Beneke:2011nf}%
  \BibitemOpen
  \bibfield  {author} {\bibinfo {author} {\bibfnamefont {M.}~\bibnamefont
  {Beneke}}\ and\ \bibinfo {author} {\bibfnamefont {J.}~\bibnamefont
  {Rohrwild}},\ }\href {https://doi.org/10.1140/epjc/s10052-011-1818-8}
  {\bibfield  {journal} {\bibinfo  {journal} {Eur. Phys. J. C}\ }\textbf
  {\bibinfo {volume} {71}},\ \bibinfo {pages} {1818} (\bibinfo {year}
  {2011})},\ \Eprint {https://arxiv.org/abs/1110.3228} {arXiv:1110.3228
  [hep-ph]} \BibitemShut {NoStop}%
\bibitem [{\citenamefont {Wang}(2016)}]{Wang:2016qii}%
  \BibitemOpen
  \bibfield  {author} {\bibinfo {author} {\bibfnamefont {Y.-M.}\ \bibnamefont
  {Wang}},\ }\href {https://doi.org/10.1007/JHEP09(2016)159} {\bibfield
  {journal} {\bibinfo  {journal} {JHEP\phantom{}}\ }\textbf {\bibinfo {volume}
  {09}},\ \bibinfo {pages} {159} (\bibinfo {year} {2016})},\ \Eprint
  {https://arxiv.org/abs/1606.03080} {arXiv:1606.03080 [hep-ph]} \BibitemShut
  {NoStop}%
\bibitem [{\citenamefont {Beneke}\ \emph {et~al.}(2018)\citenamefont {Beneke},
  \citenamefont {Braun}, \citenamefont {Ji},\ and\ \citenamefont
  {Wei}}]{Beneke:2018wjp}%
  \BibitemOpen
  \bibfield  {author} {\bibinfo {author} {\bibfnamefont {M.}~\bibnamefont
  {Beneke}}, \bibinfo {author} {\bibfnamefont {V.~M.}\ \bibnamefont {Braun}},
  \bibinfo {author} {\bibfnamefont {Y.}~\bibnamefont {Ji}},\ and\ \bibinfo
  {author} {\bibfnamefont {Y.-B.}\ \bibnamefont {Wei}},\ }\href
  {https://doi.org/10.1007/JHEP07(2018)154} {\bibfield  {journal} {\bibinfo
  {journal} {JHEP\phantom{}}\ }\textbf {\bibinfo {volume} {07}},\ \bibinfo
  {pages} {154} (\bibinfo {year} {2018})},\ \Eprint
  {https://arxiv.org/abs/1804.04962} {arXiv:1804.04962 [hep-ph]} \BibitemShut
  {NoStop}%
\bibitem [{\citenamefont {Beneke}\ \emph {et~al.}(2021)\citenamefont {Beneke},
  \citenamefont {B\"oer}, \citenamefont {Rigatos},\ and\ \citenamefont
  {Vos}}]{Beneke:2021rjf}%
  \BibitemOpen
  \bibfield  {author} {\bibinfo {author} {\bibfnamefont {M.}~\bibnamefont
  {Beneke}}, \bibinfo {author} {\bibfnamefont {P.}~\bibnamefont {B\"oer}},
  \bibinfo {author} {\bibfnamefont {P.}~\bibnamefont {Rigatos}},\ and\ \bibinfo
  {author} {\bibfnamefont {K.~K.}\ \bibnamefont {Vos}},\ }\href
  {https://doi.org/10.1140/epjc/s10052-021-09388-y} {\bibfield  {journal}
  {\bibinfo  {journal} {Eur. Phys. J. C}\ }\textbf {\bibinfo {volume} {81}},\
  \bibinfo {pages} {638} (\bibinfo {year} {2021})},\ \Eprint
  {https://arxiv.org/abs/2102.10060} {arXiv:2102.10060 [hep-ph]} \BibitemShut
  {NoStop}%
\bibitem [{\citenamefont {Ivanov}\ and\ \citenamefont
  {Melikhov}(2022{\natexlab{a}})}]{Ivanov:2021jsr}%
  \BibitemOpen
  \bibfield  {author} {\bibinfo {author} {\bibfnamefont {M.~A.}\ \bibnamefont
  {Ivanov}}\ and\ \bibinfo {author} {\bibfnamefont {D.}~\bibnamefont
  {Melikhov}},\ }\href {https://doi.org/10.1103/PhysRevD.105.014028} {\bibfield
   {journal} {\bibinfo  {journal} {Phys. Rev. D}\ }\textbf {\bibinfo {volume}
  {105}},\ \bibinfo {pages} {014028} (\bibinfo {year} {2022}{\natexlab{a}})},\
  \bibinfo {note} {[Erratum: Phys. Rev. D \textbf{106}, 119901 (2022)]},\
  \Eprint {https://arxiv.org/abs/2107.07247} {arXiv:2107.07247 [hep-ph]}
  \BibitemShut {NoStop}%
\bibitem [{\citenamefont {Albrecht}\ \emph {et~al.}(2021)\citenamefont
  {Albrecht}, \citenamefont {Stamou}, \citenamefont {Ziegler},\ and\
  \citenamefont {Zwicky}}]{Albrecht:2019zul}%
  \BibitemOpen
  \bibfield  {author} {\bibinfo {author} {\bibfnamefont {J.}~\bibnamefont
  {Albrecht}}, \bibinfo {author} {\bibfnamefont {E.}~\bibnamefont {Stamou}},
  \bibinfo {author} {\bibfnamefont {R.}~\bibnamefont {Ziegler}},\ and\ \bibinfo
  {author} {\bibfnamefont {R.}~\bibnamefont {Zwicky}},\ }\href
  {https://doi.org/10.1007/JHEP09(2021)139} {\bibfield  {journal} {\bibinfo
  {journal} {JHEP\phantom{}}\ }\textbf {\bibinfo {volume} {09}},\ \bibinfo
  {pages} {139} (\bibinfo {year} {2021})},\ \Eprint
  {https://arxiv.org/abs/1911.05018} {arXiv:1911.05018 [hep-ph]} \BibitemShut
  {NoStop}%
\bibitem [{\citenamefont {Wang}\ \emph {et~al.}(2022)\citenamefont {Wang},
  \citenamefont {Wang},\ and\ \citenamefont {Wei}}]{Wang:2021yrr}%
  \BibitemOpen
  \bibfield  {author} {\bibinfo {author} {\bibfnamefont {C.}~\bibnamefont
  {Wang}}, \bibinfo {author} {\bibfnamefont {Y.-M.}\ \bibnamefont {Wang}},\
  and\ \bibinfo {author} {\bibfnamefont {Y.-B.}\ \bibnamefont {Wei}},\ }\href
  {https://doi.org/10.1007/JHEP02(2022)141} {\bibfield  {journal} {\bibinfo
  {journal} {JHEP\phantom{}}\ }\textbf {\bibinfo {volume} {02}},\ \bibinfo
  {pages} {141} (\bibinfo {year} {2022})},\ \Eprint
  {https://arxiv.org/abs/2111.11811} {arXiv:2111.11811 [hep-ph]} \BibitemShut
  {NoStop}%
\bibitem [{\citenamefont {Colangelo}\ \emph {et~al.}(2019)\citenamefont
  {Colangelo}, \citenamefont {Hoferichter},\ and\ \citenamefont
  {Stoffer}}]{Colangelo:2018mtw}%
  \BibitemOpen
  \bibfield  {author} {\bibinfo {author} {\bibfnamefont {G.}~\bibnamefont
  {Colangelo}}, \bibinfo {author} {\bibfnamefont {M.}~\bibnamefont
  {Hoferichter}},\ and\ \bibinfo {author} {\bibfnamefont {P.}~\bibnamefont
  {Stoffer}},\ }\href {https://doi.org/10.1007/JHEP02(2019)006} {\bibfield
  {journal} {\bibinfo  {journal} {JHEP\phantom{}}\ }\textbf {\bibinfo {volume}
  {02}},\ \bibinfo {pages} {006} (\bibinfo {year} {2019})},\ \Eprint
  {https://arxiv.org/abs/1810.00007} {arXiv:1810.00007 [hep-ph]} \BibitemShut
  {NoStop}%
\bibitem [{\citenamefont {Bharucha}\ \emph {et~al.}(2016)\citenamefont
  {Bharucha}, \citenamefont {Straub},\ and\ \citenamefont
  {Zwicky}}]{Bharucha:2015bzk}%
  \BibitemOpen
  \bibfield  {author} {\bibinfo {author} {\bibfnamefont {A.}~\bibnamefont
  {Bharucha}}, \bibinfo {author} {\bibfnamefont {D.~M.}\ \bibnamefont
  {Straub}},\ and\ \bibinfo {author} {\bibfnamefont {R.}~\bibnamefont
  {Zwicky}},\ }\href {https://doi.org/10.1007/JHEP08(2016)098} {\bibfield
  {journal} {\bibinfo  {journal} {JHEP\phantom{}}\ }\textbf {\bibinfo {volume}
  {08}},\ \bibinfo {pages} {098} (\bibinfo {year} {2016})},\ \Eprint
  {https://arxiv.org/abs/1503.05534} {arXiv:1503.05534 [hep-ph]} \BibitemShut
  {NoStop}%
\bibitem [{\citenamefont {Bardeen}\ and\ \citenamefont
  {Tung}(1968)}]{Bardeen:1968ebo}%
  \BibitemOpen
  \bibfield  {author} {\bibinfo {author} {\bibfnamefont {W.~A.}\ \bibnamefont
  {Bardeen}}\ and\ \bibinfo {author} {\bibfnamefont {W.~K.}\ \bibnamefont
  {Tung}},\ }\href {https://doi.org/10.1103/PhysRev.173.1423} {\bibfield
  {journal} {\bibinfo  {journal} {Phys. Rev.}\ }\textbf {\bibinfo {volume}
  {173}},\ \bibinfo {pages} {1423} (\bibinfo {year} {1968})},\ \bibinfo {note}
  {[Erratum: Phys. Rev. D \textbf{4}, 3229 (1971)]}\BibitemShut {NoStop}%
\bibitem [{\citenamefont {Tarrach}(1975)}]{Tarrach:1975tu}%
  \BibitemOpen
  \bibfield  {author} {\bibinfo {author} {\bibfnamefont {R.}~\bibnamefont
  {Tarrach}},\ }\href {https://doi.org/10.1007/BF02894857} {\bibfield
  {journal} {\bibinfo  {journal} {Nuovo Cim. A}\ }\textbf {\bibinfo {volume}
  {28}},\ \bibinfo {pages} {409} (\bibinfo {year} {1975})}\BibitemShut
  {NoStop}%
\bibitem [{\citenamefont {Aebischer}\ \emph {et~al.}(2017)\citenamefont
  {Aebischer}, \citenamefont {Fael}, \citenamefont {Greub},\ and\ \citenamefont
  {Virto}}]{Aebischer:2017gaw}%
  \BibitemOpen
  \bibfield  {author} {\bibinfo {author} {\bibfnamefont {J.}~\bibnamefont
  {Aebischer}}, \bibinfo {author} {\bibfnamefont {M.}~\bibnamefont {Fael}},
  \bibinfo {author} {\bibfnamefont {C.}~\bibnamefont {Greub}},\ and\ \bibinfo
  {author} {\bibfnamefont {J.}~\bibnamefont {Virto}},\ }\href
  {https://doi.org/10.1007/JHEP09(2017)158} {\bibfield  {journal} {\bibinfo
  {journal} {JHEP\phantom{}}\ }\textbf {\bibinfo {volume} {09}},\ \bibinfo
  {pages} {158} (\bibinfo {year} {2017})},\ \Eprint
  {https://arxiv.org/abs/1704.06639} {arXiv:1704.06639 [hep-ph]} \BibitemShut
  {NoStop}%
\bibitem [{\citenamefont {Jenkins}\ \emph {et~al.}(2018)\citenamefont
  {Jenkins}, \citenamefont {Manohar},\ and\ \citenamefont
  {Stoffer}}]{Jenkins:2017jig}%
  \BibitemOpen
  \bibfield  {author} {\bibinfo {author} {\bibfnamefont {E.~E.}\ \bibnamefont
  {Jenkins}}, \bibinfo {author} {\bibfnamefont {A.~V.}\ \bibnamefont
  {Manohar}},\ and\ \bibinfo {author} {\bibfnamefont {P.}~\bibnamefont
  {Stoffer}},\ }\href {https://doi.org/10.1007/JHEP03(2018)016} {\bibfield
  {journal} {\bibinfo  {journal} {JHEP\phantom{}}\ }\textbf {\bibinfo {volume}
  {03}},\ \bibinfo {pages} {016} (\bibinfo {year} {2018})},\ \Eprint
  {https://arxiv.org/abs/1709.04486} {arXiv:1709.04486 [hep-ph]} \BibitemShut
  {NoStop}%
\bibitem [{\citenamefont {Khodjamirian}\ and\ \citenamefont
  {Wyler}(2001)}]{Khodjamirian:2001ga}%
  \BibitemOpen
  \bibfield  {author} {\bibinfo {author} {\bibfnamefont {A.}~\bibnamefont
  {Khodjamirian}}\ and\ \bibinfo {author} {\bibfnamefont {D.}~\bibnamefont
  {Wyler}}\ }\href {https://doi.org/10.1142/9789812777478\_0014}
  {10.1142/9789812777478\_0014} (\bibinfo {year} {2001}),\ \Eprint
  {https://arxiv.org/abs/hep-ph/0111249} {arXiv:hep-ph/0111249} \BibitemShut
  {NoStop}%
\bibitem [{\citenamefont {Janowski}\ \emph {et~al.}(2021)\citenamefont
  {Janowski}, \citenamefont {Pullin},\ and\ \citenamefont
  {Zwicky}}]{Janowski:2021yvz}%
  \BibitemOpen
  \bibfield  {author} {\bibinfo {author} {\bibfnamefont {T.}~\bibnamefont
  {Janowski}}, \bibinfo {author} {\bibfnamefont {B.}~\bibnamefont {Pullin}},\
  and\ \bibinfo {author} {\bibfnamefont {R.}~\bibnamefont {Zwicky}},\ }\href
  {https://doi.org/10.1007/JHEP12(2021)008} {\bibfield  {journal} {\bibinfo
  {journal} {JHEP\phantom{}}\ }\textbf {\bibinfo {volume} {12}},\ \bibinfo
  {pages} {008} (\bibinfo {year} {2021})},\ \Eprint
  {https://arxiv.org/abs/2106.13616} {arXiv:2106.13616 [hep-ph]} \BibitemShut
  {NoStop}%
\bibitem [{\citenamefont {Bijnens}\ \emph {et~al.}(1993)\citenamefont
  {Bijnens}, \citenamefont {Ecker},\ and\ \citenamefont
  {Gasser}}]{Bijnens:1992en}%
  \BibitemOpen
  \bibfield  {author} {\bibinfo {author} {\bibfnamefont {J.}~\bibnamefont
  {Bijnens}}, \bibinfo {author} {\bibfnamefont {G.}~\bibnamefont {Ecker}},\
  and\ \bibinfo {author} {\bibfnamefont {J.}~\bibnamefont {Gasser}},\ }\href
  {https://doi.org/10.1016/0550-3213(93)90259-R} {\bibfield  {journal}
  {\bibinfo  {journal} {Nucl. Phys. B}\ }\textbf {\bibinfo {volume} {396}},\
  \bibinfo {pages} {81} (\bibinfo {year} {1993})},\ \Eprint
  {https://arxiv.org/abs/hep-ph/9209261} {arXiv:hep-ph/9209261} \BibitemShut
  {NoStop}%
\bibitem [{\citenamefont {Bijnens}\ \emph {et~al.}(1994)\citenamefont
  {Bijnens}, \citenamefont {Colangelo}, \citenamefont {Ecker},\ and\
  \citenamefont {Gasser}}]{Bijnens:1994me}%
  \BibitemOpen
  \bibfield  {author} {\bibinfo {author} {\bibfnamefont {J.}~\bibnamefont
  {Bijnens}}, \bibinfo {author} {\bibfnamefont {G.}~\bibnamefont {Colangelo}},
  \bibinfo {author} {\bibfnamefont {G.}~\bibnamefont {Ecker}},\ and\ \bibinfo
  {author} {\bibfnamefont {J.}~\bibnamefont {Gasser}}\ }(\bibinfo {year}
  {1994})\ \Eprint {https://arxiv.org/abs/hep-ph/9411311}
  {arXiv:hep-ph/9411311} \BibitemShut {NoStop}%
\bibitem [{\citenamefont {Pal}(2007)}]{Pal:2007dc}%
  \BibitemOpen
  \bibfield  {author} {\bibinfo {author} {\bibfnamefont {P.~B.}\ \bibnamefont
  {Pal}}\ }(\bibinfo {year} {2007})\ \Eprint
  {https://arxiv.org/abs/physics/0703214} {arXiv:physics/0703214} \BibitemShut
  {NoStop}%
\bibitem [{\citenamefont {Ivanov}\ and\ \citenamefont
  {Melikhov}(2022{\natexlab{b}})}]{Ivanov:2022uum}%
  \BibitemOpen
  \bibfield  {author} {\bibinfo {author} {\bibfnamefont {M.~A.}\ \bibnamefont
  {Ivanov}}\ and\ \bibinfo {author} {\bibfnamefont {D.}~\bibnamefont
  {Melikhov}},\ }\href {https://doi.org/10.1103/PhysRevD.105.094038} {\bibfield
   {journal} {\bibinfo  {journal} {Phys. Rev. D}\ }\textbf {\bibinfo {volume}
  {105}},\ \bibinfo {pages} {094038} (\bibinfo {year} {2022}{\natexlab{b}})},\
  \Eprint {https://arxiv.org/abs/2204.02792} {arXiv:2204.02792 [hep-ph]}
  \BibitemShut {NoStop}%
\bibitem [{\citenamefont {Bardin}\ and\ \citenamefont
  {Ivanov}(1976)}]{Bardin:1976wv}%
  \BibitemOpen
  \bibfield  {author} {\bibinfo {author} {\bibfnamefont {D.~Y.}\ \bibnamefont
  {Bardin}}\ and\ \bibinfo {author} {\bibfnamefont {E.~A.}\ \bibnamefont
  {Ivanov}},\ }\href@noop {} {\bibfield  {journal} {\bibinfo  {journal} {Sov.
  J. Part. Nucl.}\ }\textbf {\bibinfo {volume} {7}},\ \bibinfo {pages} {286}
  (\bibinfo {year} {1976})}\BibitemShut {NoStop}%
\bibitem [{\citenamefont {Okubo}(1963)}]{Okubo:1963fa}%
  \BibitemOpen
  \bibfield  {author} {\bibinfo {author} {\bibfnamefont {S.}~\bibnamefont
  {Okubo}},\ }\href {https://doi.org/10.1016/S0375-9601(63)92548-9} {\bibfield
  {journal} {\bibinfo  {journal} {Phys. Lett.}\ }\textbf {\bibinfo {volume}
  {5}},\ \bibinfo {pages} {165} (\bibinfo {year} {1963})}\BibitemShut {NoStop}%
\bibitem [{\citenamefont {Zweig}(1964)}]{Zweig:1964jf}%
  \BibitemOpen
  \bibfield  {author} {\bibinfo {author} {\bibfnamefont {G.}~\bibnamefont
  {Zweig}},\ }\bibinfo {title} {{An $SU(3)$ model for strong interaction
  symmetry and its breaking. Version 2}},\ in\ \href@noop {} {\emph {\bibinfo
  {booktitle} {{Developments in the quark theory of hadrons. Vol.~1.
  1964--1978}}}},\ \bibinfo {editor} {edited by\ \bibinfo {editor}
  {\bibfnamefont {D.~B.}\ \bibnamefont {Lichtenberg}}\ and\ \bibinfo {editor}
  {\bibfnamefont {S.~P.}\ \bibnamefont {Rosen}}}\ (\bibinfo {year} {1964})\
  pp.\ \bibinfo {pages} {22--101}\BibitemShut {NoStop}%
\bibitem [{\citenamefont {Iizuka}(1966)}]{Iizuka:1966fk}%
  \BibitemOpen
  \bibfield  {author} {\bibinfo {author} {\bibfnamefont {J.}~\bibnamefont
  {Iizuka}},\ }\href {https://doi.org/10.1143/PTPS.37.21} {\bibfield  {journal}
  {\bibinfo  {journal} {Prog. Theor. Phys. Suppl.}\ }\textbf {\bibinfo {volume}
  {37}},\ \bibinfo {pages} {21} (\bibinfo {year} {1966})}\BibitemShut {NoStop}%
\bibitem [{\citenamefont {Colangelo}\ and\ \citenamefont
  {Khodjamirian}(2000)}]{Colangelo:2000dp}%
  \BibitemOpen
  \bibfield  {author} {\bibinfo {author} {\bibfnamefont {P.}~\bibnamefont
  {Colangelo}}\ and\ \bibinfo {author} {\bibfnamefont {A.}~\bibnamefont
  {Khodjamirian}},\ }\bibinfo {title} {{QCD sum rules, a modern perspective}},\
  in\ \href {https://doi.org/10.1142/9789812810458_0033} {\emph {\bibinfo
  {booktitle} {{At The Frontier of Particle Physics}}}},\ \bibinfo {editor}
  {edited by\ \bibinfo {editor} {\bibfnamefont {M.}~\bibnamefont {Shifman}}\
  and\ \bibinfo {editor} {\bibfnamefont {B.}~\bibnamefont {Ioffe}}}\ (\bibinfo
  {year} {2000})\ pp.\ \bibinfo {pages} {1495--1576},\ \Eprint
  {https://arxiv.org/abs/hep-ph/0010175} {arXiv:hep-ph/0010175} \BibitemShut
  {NoStop}%
\bibitem [{\citenamefont {Khodjamirian}(2020)}]{Khodjamirian:2020btr}%
  \BibitemOpen
  \bibfield  {author} {\bibinfo {author} {\bibfnamefont {A.}~\bibnamefont
  {Khodjamirian}},\ }\href@noop {} {\emph {\bibinfo {title} {{Hadron Form
  Factors}: {From Basic Phenomenology to QCD Sum Rules}}}}\ (\bibinfo
  {publisher} {CRC Press, Taylor \& Francis Group},\ \bibinfo {address} {Boca
  Raton, FL, USA},\ \bibinfo {year} {2020})\BibitemShut {NoStop}%
\bibitem [{\citenamefont {Horgan}\ \emph {et~al.}(2014)\citenamefont {Horgan},
  \citenamefont {Liu}, \citenamefont {Meinel},\ and\ \citenamefont
  {Wingate}}]{Horgan:2013hoa}%
  \BibitemOpen
  \bibfield  {author} {\bibinfo {author} {\bibfnamefont {R.~R.}\ \bibnamefont
  {Horgan}}, \bibinfo {author} {\bibfnamefont {Z.}~\bibnamefont {Liu}},
  \bibinfo {author} {\bibfnamefont {S.}~\bibnamefont {Meinel}},\ and\ \bibinfo
  {author} {\bibfnamefont {M.}~\bibnamefont {Wingate}},\ }\href
  {https://doi.org/10.1103/PhysRevD.89.094501} {\bibfield  {journal} {\bibinfo
  {journal} {Phys. Rev. D}\ }\textbf {\bibinfo {volume} {89}},\ \bibinfo
  {pages} {094501} (\bibinfo {year} {2014})},\ \Eprint
  {https://arxiv.org/abs/1310.3722} {arXiv:1310.3722 [hep-lat]} \BibitemShut
  {NoStop}%
\bibitem [{\citenamefont {Zanke}\ \emph {et~al.}(2021)\citenamefont {Zanke},
  \citenamefont {Hoferichter},\ and\ \citenamefont {Kubis}}]{Zanke:2021wiq}%
  \BibitemOpen
  \bibfield  {author} {\bibinfo {author} {\bibfnamefont {M.}~\bibnamefont
  {Zanke}}, \bibinfo {author} {\bibfnamefont {M.}~\bibnamefont {Hoferichter}},\
  and\ \bibinfo {author} {\bibfnamefont {B.}~\bibnamefont {Kubis}},\ }\href
  {https://doi.org/10.1007/JHEP07(2021)106} {\bibfield  {journal} {\bibinfo
  {journal} {JHEP\phantom{}}\ }\textbf {\bibinfo {volume} {07}},\ \bibinfo
  {pages} {106} (\bibinfo {year} {2021})},\ \Eprint
  {https://arxiv.org/abs/2103.09829} {arXiv:2103.09829 [hep-ph]} \BibitemShut
  {NoStop}%
\bibitem [{\citenamefont {Zyla}\ \emph {et~al.}(2020)\citenamefont {Zyla} \emph
  {et~al.}}]{ParticleDataGroup:2020ssz}%
  \BibitemOpen
  \bibfield  {author} {\bibinfo {author} {\bibfnamefont {P.~A.}\ \bibnamefont
  {Zyla}} \emph {et~al.} (\bibinfo {collaboration} {Particle Data Group}),\
  }\href {https://doi.org/10.1093/ptep/ptaa104} {\bibfield  {journal} {\bibinfo
   {journal} {PTEP}\ }\textbf {\bibinfo {volume} {2020}},\ \bibinfo {pages}
  {083C01} (\bibinfo {year} {2020})}\BibitemShut {NoStop}%
\bibitem [{\citenamefont {Daub}\ \emph {et~al.}(2016)\citenamefont {Daub},
  \citenamefont {Hanhart},\ and\ \citenamefont {Kubis}}]{Daub:2015xja}%
  \BibitemOpen
  \bibfield  {author} {\bibinfo {author} {\bibfnamefont {J.~T.}\ \bibnamefont
  {Daub}}, \bibinfo {author} {\bibfnamefont {C.}~\bibnamefont {Hanhart}},\ and\
  \bibinfo {author} {\bibfnamefont {B.}~\bibnamefont {Kubis}},\ }\href
  {https://doi.org/10.1007/JHEP02(2016)009} {\bibfield  {journal} {\bibinfo
  {journal} {JHEP\phantom{}}\ }\textbf {\bibinfo {volume} {02}},\ \bibinfo
  {pages} {009} (\bibinfo {year} {2016})},\ \Eprint
  {https://arxiv.org/abs/1508.06841} {arXiv:1508.06841 [hep-ph]} \BibitemShut
  {NoStop}%
\bibitem [{\citenamefont {Ropertz}\ \emph {et~al.}(2018)\citenamefont
  {Ropertz}, \citenamefont {Hanhart},\ and\ \citenamefont
  {Kubis}}]{Ropertz:2018stk}%
  \BibitemOpen
  \bibfield  {author} {\bibinfo {author} {\bibfnamefont {S.}~\bibnamefont
  {Ropertz}}, \bibinfo {author} {\bibfnamefont {C.}~\bibnamefont {Hanhart}},\
  and\ \bibinfo {author} {\bibfnamefont {B.}~\bibnamefont {Kubis}},\ }\href
  {https://doi.org/10.1140/epjc/s10052-018-6416-6} {\bibfield  {journal}
  {\bibinfo  {journal} {Eur. Phys. J. C}\ }\textbf {\bibinfo {volume} {78}},\
  \bibinfo {pages} {1000} (\bibinfo {year} {2018})},\ \Eprint
  {https://arxiv.org/abs/1809.06867} {arXiv:1809.06867 [hep-ph]} \BibitemShut
  {NoStop}%
\bibitem [{\citenamefont {Kang}\ \emph {et~al.}(2014)\citenamefont {Kang},
  \citenamefont {Kubis}, \citenamefont {Hanhart},\ and\ \citenamefont
  {Mei\ss{}ner}}]{Kang:2013jaa}%
  \BibitemOpen
  \bibfield  {author} {\bibinfo {author} {\bibfnamefont {X.-W.}\ \bibnamefont
  {Kang}}, \bibinfo {author} {\bibfnamefont {B.}~\bibnamefont {Kubis}},
  \bibinfo {author} {\bibfnamefont {C.}~\bibnamefont {Hanhart}},\ and\ \bibinfo
  {author} {\bibfnamefont {U.-G.}\ \bibnamefont {Mei\ss{}ner}},\ }\href
  {https://doi.org/10.1103/PhysRevD.89.053015} {\bibfield  {journal} {\bibinfo
  {journal} {Phys. Rev. D}\ }\textbf {\bibinfo {volume} {89}},\ \bibinfo
  {pages} {053015} (\bibinfo {year} {2014})},\ \Eprint
  {https://arxiv.org/abs/1312.1193} {arXiv:1312.1193 [hep-ph]} \BibitemShut
  {NoStop}%
\bibitem [{\citenamefont {Hanhart}\ \emph {et~al.}(2013)\citenamefont
  {Hanhart}, \citenamefont {Kup\'s\'c}, \citenamefont {Mei\ss{}ner},
  \citenamefont {Stollenwerk},\ and\ \citenamefont {Wirzba}}]{Hanhart:2013vba}%
  \BibitemOpen
  \bibfield  {author} {\bibinfo {author} {\bibfnamefont {C.}~\bibnamefont
  {Hanhart}}, \bibinfo {author} {\bibfnamefont {A.}~\bibnamefont {Kup\'s\'c}},
  \bibinfo {author} {\bibfnamefont {U.-G.}\ \bibnamefont {Mei\ss{}ner}},
  \bibinfo {author} {\bibfnamefont {F.}~\bibnamefont {Stollenwerk}},\ and\
  \bibinfo {author} {\bibfnamefont {A.}~\bibnamefont {Wirzba}},\ }\href
  {https://doi.org/10.1140/epjc/s10052-013-2668-3} {\bibfield  {journal}
  {\bibinfo  {journal} {Eur. Phys. J. C}\ }\textbf {\bibinfo {volume} {73}},\
  \bibinfo {pages} {2668} (\bibinfo {year} {2013})},\ \bibinfo {note}
  {[Erratum: Eur. Phys. J. C \textbf{75}, 242 (2015)]},\ \Eprint
  {https://arxiv.org/abs/1307.5654} {arXiv:1307.5654 [hep-ph]} \BibitemShut
  {NoStop}%
\bibitem [{\citenamefont {Holz}\ \emph {et~al.}(2022)\citenamefont {Holz},
  \citenamefont {Hanhart}, \citenamefont {Hoferichter},\ and\ \citenamefont
  {Kubis}}]{Holz:2022hwz}%
  \BibitemOpen
  \bibfield  {author} {\bibinfo {author} {\bibfnamefont {S.}~\bibnamefont
  {Holz}}, \bibinfo {author} {\bibfnamefont {C.}~\bibnamefont {Hanhart}},
  \bibinfo {author} {\bibfnamefont {M.}~\bibnamefont {Hoferichter}},\ and\
  \bibinfo {author} {\bibfnamefont {B.}~\bibnamefont {Kubis}},\ }\href
  {https://doi.org/10.1140/epjc/s10052-022-10247-7} {\bibfield  {journal}
  {\bibinfo  {journal} {Eur. Phys. J. C}\ }\textbf {\bibinfo {volume} {82}},\
  \bibinfo {pages} {434} (\bibinfo {year} {2022})},\ \bibinfo {note}
  {[Addendum: Eur. Phys. J. C \textbf{82}, 1159 (2022)]},\ \Eprint
  {https://arxiv.org/abs/2202.05846} {arXiv:2202.05846 [hep-ph]} \BibitemShut
  {NoStop}%
\bibitem [{\citenamefont {Mertig}\ \emph {et~al.}(1991)\citenamefont {Mertig},
  \citenamefont {Böhm},\ and\ \citenamefont {Denner}}]{Mertig:1990an}%
  \BibitemOpen
  \bibfield  {author} {\bibinfo {author} {\bibfnamefont {R.}~\bibnamefont
  {Mertig}}, \bibinfo {author} {\bibfnamefont {M.}~\bibnamefont {Böhm}},\ and\
  \bibinfo {author} {\bibfnamefont {A.}~\bibnamefont {Denner}},\ }\href
  {https://doi.org/10.1016/0010-4655(91)90130-D} {\bibfield  {journal}
  {\bibinfo  {journal} {Comput. Phys. Commun.}\ }\textbf {\bibinfo {volume}
  {64}},\ \bibinfo {pages} {345} (\bibinfo {year} {1991})}\BibitemShut
  {NoStop}%
\bibitem [{\citenamefont {Shtabovenko}\ \emph {et~al.}(2016)\citenamefont
  {Shtabovenko}, \citenamefont {Mertig},\ and\ \citenamefont
  {Orellana}}]{Shtabovenko:2016sxi}%
  \BibitemOpen
  \bibfield  {author} {\bibinfo {author} {\bibfnamefont {V.}~\bibnamefont
  {Shtabovenko}}, \bibinfo {author} {\bibfnamefont {R.}~\bibnamefont
  {Mertig}},\ and\ \bibinfo {author} {\bibfnamefont {F.}~\bibnamefont
  {Orellana}},\ }\href {https://doi.org/10.1016/j.cpc.2016.06.008} {\bibfield
  {journal} {\bibinfo  {journal} {Comput. Phys. Commun.}\ }\textbf {\bibinfo
  {volume} {207}},\ \bibinfo {pages} {432} (\bibinfo {year} {2016})},\ \Eprint
  {https://arxiv.org/abs/1601.01167} {arXiv:1601.01167 [hep-ph]} \BibitemShut
  {NoStop}%
\bibitem [{\citenamefont {Shtabovenko}\ \emph {et~al.}(2020)\citenamefont
  {Shtabovenko}, \citenamefont {Mertig},\ and\ \citenamefont
  {Orellana}}]{Shtabovenko:2020gxv}%
  \BibitemOpen
  \bibfield  {author} {\bibinfo {author} {\bibfnamefont {V.}~\bibnamefont
  {Shtabovenko}}, \bibinfo {author} {\bibfnamefont {R.}~\bibnamefont
  {Mertig}},\ and\ \bibinfo {author} {\bibfnamefont {F.}~\bibnamefont
  {Orellana}},\ }\href {https://doi.org/10.1016/j.cpc.2020.107478} {\bibfield
  {journal} {\bibinfo  {journal} {Comput. Phys. Commun.}\ }\textbf {\bibinfo
  {volume} {256}},\ \bibinfo {pages} {107478} (\bibinfo {year} {2020})},\
  \Eprint {https://arxiv.org/abs/2001.04407} {arXiv:2001.04407 [hep-ph]}
  \BibitemShut {NoStop}%
\bibitem [{\citenamefont {Patel}(2015)}]{Patel:2015tea}%
  \BibitemOpen
  \bibfield  {author} {\bibinfo {author} {\bibfnamefont {H.~H.}\ \bibnamefont
  {Patel}},\ }\href {https://doi.org/10.1016/j.cpc.2015.08.017} {\bibfield
  {journal} {\bibinfo  {journal} {Comput. Phys. Commun.}\ }\textbf {\bibinfo
  {volume} {197}},\ \bibinfo {pages} {276} (\bibinfo {year} {2015})},\ \Eprint
  {https://arxiv.org/abs/1503.01469} {arXiv:1503.01469 [hep-ph]} \BibitemShut
  {NoStop}%
\bibitem [{\citenamefont {Aoki}\ \emph {et~al.}(2022)\citenamefont {Aoki} \emph
  {et~al.}}]{Aoki:2021kgd}%
  \BibitemOpen
  \bibfield  {author} {\bibinfo {author} {\bibfnamefont {Y.}~\bibnamefont
  {Aoki}} \emph {et~al.} (\bibinfo {collaboration} {Flavour Lattice Averaging
  Group (FLAG)}),\ }\href {https://doi.org/10.1140/epjc/s10052-022-10536-1}
  {\bibfield  {journal} {\bibinfo  {journal} {Eur. Phys. J. C}\ }\textbf
  {\bibinfo {volume} {82}},\ \bibinfo {pages} {869} (\bibinfo {year} {2022})},\
  \Eprint {https://arxiv.org/abs/2111.09849} {arXiv:2111.09849 [hep-lat]}
  \BibitemShut {NoStop}%
\bibitem [{\citenamefont {Bazavov}\ \emph {et~al.}(2018)\citenamefont {Bazavov}
  \emph {et~al.}}]{TUMQCD:2018fsq}%
  \BibitemOpen
  \bibfield  {author} {\bibinfo {author} {\bibfnamefont {A.}~\bibnamefont
  {Bazavov}} \emph {et~al.} (\bibinfo {collaboration} {TUMQCD, Fermilab
  Lattice, MILC}),\ }in\ \href@noop {} {\emph {\bibinfo {booktitle} {{13th
  Conference on the Intersections of Particle and Nuclear Physics}}}}\
  (\bibinfo {year} {2018})\ \Eprint {https://arxiv.org/abs/1810.00250}
  {arXiv:1810.00250 [hep-lat]} \BibitemShut {NoStop}%
\bibitem [{\citenamefont {Bussone}\ \emph {et~al.}(2016)\citenamefont {Bussone}
  \emph {et~al.}}]{ETM:2016nbo}%
  \BibitemOpen
  \bibfield  {author} {\bibinfo {author} {\bibfnamefont {A.}~\bibnamefont
  {Bussone}} \emph {et~al.} (\bibinfo {collaboration} {ETM}),\ }\href
  {https://doi.org/10.1103/PhysRevD.93.114505} {\bibfield  {journal} {\bibinfo
  {journal} {Phys. Rev. D}\ }\textbf {\bibinfo {volume} {93}},\ \bibinfo
  {pages} {114505} (\bibinfo {year} {2016})},\ \Eprint
  {https://arxiv.org/abs/1603.04306} {arXiv:1603.04306 [hep-lat]} \BibitemShut
  {NoStop}%
\bibitem [{\citenamefont {Dowdall}\ \emph {et~al.}(2013)\citenamefont
  {Dowdall}, \citenamefont {Davies}, \citenamefont {Horgan}, \citenamefont
  {Monahan},\ and\ \citenamefont {Shigemitsu}}]{Dowdall:2013tga}%
  \BibitemOpen
  \bibfield  {author} {\bibinfo {author} {\bibfnamefont {R.~J.}\ \bibnamefont
  {Dowdall}}, \bibinfo {author} {\bibfnamefont {C.~T.~H.}\ \bibnamefont
  {Davies}}, \bibinfo {author} {\bibfnamefont {R.~R.}\ \bibnamefont {Horgan}},
  \bibinfo {author} {\bibfnamefont {C.~J.}\ \bibnamefont {Monahan}},\ and\
  \bibinfo {author} {\bibfnamefont {J.}~\bibnamefont {Shigemitsu}} (\bibinfo
  {collaboration} {HPQCD}),\ }\href
  {https://doi.org/10.1103/PhysRevLett.110.222003} {\bibfield  {journal}
  {\bibinfo  {journal} {Phys. Rev. Lett.}\ }\textbf {\bibinfo {volume} {110}},\
  \bibinfo {pages} {222003} (\bibinfo {year} {2013})},\ \Eprint
  {https://arxiv.org/abs/1302.2644} {arXiv:1302.2644 [hep-lat]} \BibitemShut
  {NoStop}%
\bibitem [{\citenamefont {Hughes}\ \emph {et~al.}(2018)\citenamefont {Hughes},
  \citenamefont {Davies},\ and\ \citenamefont {Monahan}}]{Hughes:2017spc}%
  \BibitemOpen
  \bibfield  {author} {\bibinfo {author} {\bibfnamefont {C.}~\bibnamefont
  {Hughes}}, \bibinfo {author} {\bibfnamefont {C.~T.~H.}\ \bibnamefont
  {Davies}},\ and\ \bibinfo {author} {\bibfnamefont {C.~J.}\ \bibnamefont
  {Monahan}},\ }\href {https://doi.org/10.1103/PhysRevD.97.054509} {\bibfield
  {journal} {\bibinfo  {journal} {Phys. Rev. D}\ }\textbf {\bibinfo {volume}
  {97}},\ \bibinfo {pages} {054509} (\bibinfo {year} {2018})},\ \Eprint
  {https://arxiv.org/abs/1711.09981} {arXiv:1711.09981 [hep-lat]} \BibitemShut
  {NoStop}%
\bibitem [{\citenamefont {Leljak}\ \emph {et~al.}(2021)\citenamefont {Leljak},
  \citenamefont {Meli\'c},\ and\ \citenamefont {van Dyk}}]{Leljak:2021vte}%
  \BibitemOpen
  \bibfield  {author} {\bibinfo {author} {\bibfnamefont {D.}~\bibnamefont
  {Leljak}}, \bibinfo {author} {\bibfnamefont {B.}~\bibnamefont {Meli\'c}},\
  and\ \bibinfo {author} {\bibfnamefont {D.}~\bibnamefont {van Dyk}},\ }\href
  {https://doi.org/10.1007/JHEP07(2021)036} {\bibfield  {journal} {\bibinfo
  {journal} {JHEP\phantom{}}\ }\textbf {\bibinfo {volume} {07}},\ \bibinfo
  {pages} {036} (\bibinfo {year} {2021})},\ \Eprint
  {https://arxiv.org/abs/2102.07233} {arXiv:2102.07233 [hep-ph]} \BibitemShut
  {NoStop}%
\end{thebibliography}%

\end{document}